\documentclass[12pt]{article}
\usepackage{fullpage}
\usepackage{times}

\ifx\pdftexversion\undefined
  \usepackage[dvips]{graphics}
\else
  \usepackage[pdftex]{graphics}
\fi

\usepackage{xspace}
\usepackage{latexsym,amssymb}
\usepackage{epsfig}
\usepackage{math-cmds}
\usepackage{verbatim}
\usepackage{proof}
\usepackage{code}
\usepackage{math-envs}

\usepackage{amsmath}
\usepackage{rotating}
\usepackage{tabularx}
\usepackage[
pageanchor=true,
plainpages=false,
pdfpagelabels, % makes the status line page labels the same as latex ones
bookmarks,
bookmarksnumbered,
pdfborder=0 0 0,
pdfpagemode=UseOutlines]{hyperref}
\usepackage{subfigure}

\newcommand{\AFL}{\textsf{AFL}\xspace}

\newcommand{\afl}{\AFL}

\newcommand{\aml}{\textsf{AML}\xspace}

\renewcommand{\paragraph}[1]{{\bf {#1}}}
\newcommand{\todo}[1]{}

\newcommand{\rlabel}[1]{\hspace*{-1mm}\mbox{\small{\bf ({#1})}}}

\newcommand{\out}[1] {}

\newcounter{codeLineCntr}

\newcommand{\fixedCodeFrame}[1]
{
\vspace*{-4mm}
\begin{center}
\fbox{
\vspace*{-2mm}
\parbox[t]{0.95\columnwidth}{
\vspace*{-2mm}
#1
}
}\end{center}
\vspace*{-2mm}
}

\reversemarginpar
\newif\ifnotes
\notestrue

\newcommand{\punt}[1]{}

\newcommand{\secref}[1]{Section~\ref{sec:#1}}

\newcommand{\appref}[1]{Appendix~\ref{app:#1}}
\newcommand{\figref}[1]{Figure~\ref{fig:#1}}
\newcommand{\figreftwo}[2]{Figures \ref{fig:#1} and~\ref{fig:#2}}

\newcommand{\thmref}[1]{Theorem~\ref{thm:#1}}

\newcommand{\lemref}[1]{Lemma~\ref{lem:#1}}
\newcommand{\lemreftwo}[2]{Lemmas \ref{lem:#1} and~\ref{lem:#2}}

\newcommand{\proc}[1]{\ifmmode\mbox{\textsc{#1}}\else\textsc{#1}\fi}

  \newcommand{\func}[1]{\ifmmode\mathrm{#1}\else\textrm{#1}fi} %
\newcommand{\by}{\ensuremath{\mbox{by~}}}

\newcommand{\changeable}{\mathcd{c}}
\newcommand{\stable}{\mathcd{s}}

\newcommand{\cdparens}[1]{\mathcd{(}{#1}\mathcd{)}}

\newcommand{\Not}{\mathcd{not}}

\newcommand{\Numeral}[1]{#1}
\newcommand{\Plus}{\mathcd{+}}
\newcommand{\Minus}{\mathcd{-}}

\newcommand{\Equal}{\mathcd{=}}
\newcommand{\Less}{\mathcd{<}}

\newcommand{\unit}{\mathcd{()}}

\newcommand{\Case}{\ensuremath{\mathcd{case}}\xspace}
\newcommand{\Inl}{\ensuremath{\mathcd{in}_{\mathcd{l}}}}
\newcommand{\Inr}{\ensuremath{\mathcd{in}_{\mathcd{r}}}}
\newcommand{\inl}[1]{\Inl~{#1}}
\newcommand{\inr}[1]{\Inr~{#1}}
\newcommand{\inlr}[1]{{\mathcd{in}_{\{\mathcd{l,r}\}}}~{#1}}
\newcommand{\casearrow}{\Rightarrow}

\newcommand{\caseofflat}[5]{
\ensuremath{\Case\,{#1}\,\mathcd{of}\,\Inl\,\cdparens{#2}\,\casearrow{}\,{#3}\,\mathcd{|}\,\Inr\,\cdparens{#4}\,\casearrow\,{#5}\,\mathcd{end}}}

\newcommand{\caseof}[5]{
{\mbox{
$
\left(~
\setlength{\arraycolsep}{-1ex}
\begin{array}{rl}
\Case~{#1}~\mathcd{of}~~&~~\Inl\,\cdparens{#2}\,\casearrow{}\,{#3} \\
                  \mathcd{|}~~~&~~\Inr\,\cdparens{#4}\,\casearrow\,{#5}
\end{array}
~\right)
$
}
}}

\newcommand{\Let}{\mathcd{let}\xspace}

\newcommand{\letin}[3]{\mathcd{let}~{#1}={#2}~\mathcd{in}~{#3}}
\newcommand{\letpin}[4]{\mathcd{let}\,{#1}{\cross}{#2}\,=\,{#3}\,\mathcd{in}\,{#4}}

\newcommand{\Mod}{\mathcd{mod}\xspace}
\renewcommand{\mod}[1]{\mathcd{mod}{~\ensuremath{#1}\xspace}}
\newcommand{\Read}{\mathcd{read}\xspace}
\newcommand{\ReadIn}{\mathcd{read}\xspace}
\newcommand{\readin}[3]{\ensuremath{\mathcd{read}~{#1}~\mathcd{as}~{#2}~\mathcd{in}~{#3}}}
\newcommand{\Effect}{\mathcd{write}\xspace}
\newcommand{\effect}[1]{\mathcd{write}\cdparens{\ensuremath{#1}}}

\newcommand{\cfun}[3]{\ensuremath{\mathcd{fun}_{\changeable{}}~{#1}{(#2)}~\mathcd{is}~{#3}}}
\newcommand{\sfun}[3]{\ensuremath{\mathcd{fun}_{\stable{}}~{#1}{(#2)}~\mathcd{is}~{#3}}}
\newcommand{\scfun}[3]{\ensuremath{\mathcd{fun}_{\{\stable,\changeable\}}~{#1}{(#2)}~\mathcd{is}~{#3}}}

\newcommand{\oper}{o}
\newcommand{\op}[1]{o\cdparens{#1}}

\newcommand{\Apply}{\mathcd{apply}}

\newcommand{\capply}[2]{\ensuremath{\mathcd{apply}_{\changeable{}} \cdparens{#1,#2}}}

\newcommand{\sapply}[2]{\ensuremath{\mathcd{apply}_{\stable}} \cdparens{{#1},{#2}}}
\newcommand{\scapply}[2]{\ensuremath{\mathcd{apply}_{\{\stable,\changeable\}}} \cdparens{{#1},{#2}}}

\newcommand{\Memo}{\ensuremath{\mathcd{memo}}\xspace}
\newcommand{\smemo}[1]{\ensuremath{\mathcd{memo}_{\stable{}}~{#1}}}
\newcommand{\cmemo}[1]{\ensuremath{\mathcd{memo}_{\changeable{}}~{#1}}}
\newcommand{\scmemo}[1]{\ensuremath{\mathcd{memo}_{\{\stable,\changeable\}}~{#1}}}

\newcommand{\OpApply}[2]{\mathcd{app(\ensuremath{#1,#2})}}

\newcommand{\la}{\leftarrow}

\newcommand{\assign}{\ensuremath{\leftarrow}}

\newcommand{\tr}{\mathcd{T}}
\newcommand{\trc}{\ensuremath{\tr_c}}
\newcommand{\tri}{\ensuremath{\tr_1}}
\newcommand{\trii}{\ensuremath{\tr_2}}
\newcommand{\trs}{\ensuremath{\tr_s}}

\newcommand{\et}{\ensuremath{\varepsilon}} 
\newcommand{\treffect}[1]{\ensuremath{\Effect~{#1}}}
\newcommand{\trmod}[2]{\ensuremath{\Mod~{#1} \leftarrow {#2}}}

\newcommand{\trlet}[2]{\ensuremath{\Let~{#1}~{#2}}}
\newcommand{\trread}[5]{\ensuremath{\Read_{{{#1}} \rightarrow {#2}={#3}.{#4}}~{#5}}}

\newcommand{\store}{\ensuremath{\sigma}}
\newcommand{\s}{\store}
\newcommand{\si}{\store_1}
\newcommand{\sii}{\store_2}

\newcommand{\dom}[1]{\mathcd{dom}(#1)}

\newcommand{\gen}[1]{\mathcd{alloc}\left(#1\right)}
\newcommand{\alloc}[1]{\ensuremath{\mathcd{alloc}\left(#1\right)}}
\newcommand{\reach}[2]{\ensuremath{\mathcd{reach}\left({#1},{#2}\right)}}
\newcommand{\lift}[2]{\ensuremath{{#1}\!\uparrow\!{#2}}}

\newcommand{\wfarrow}{\stackrel{\mbox{$\mathcd{wf}$}}{\longrightarrow}}
\newcommand{\wf}[3]{\ensuremath{{#1},{#2} \wfarrow {#3}}}

\newcommand{\wfl}[4]{\ensuremath{{#1},{#2} \wfarrow {#3},{#4}}}

\newcommand{\creduces}{~\Downarrow^{\changeable}~}
\newcommand{\cnreduces}{{~\Downarrow^{\changeable}_{\emptyset}}~}
\newcommand{\cvreduces}{{~\Downarrow^{\changeable}_{\mathrm{ok}}}~}
\newcommand{\cnvreduces}{{~\Downarrow^{\changeable}_{\emptyset,\mathrm{ok}}}~}
\newcommand{\cpreduces}{{~\Downarrow^{\changeable}_{\mathrm{det}}}~}
\newcommand{\rc}{\creduces}
\newcommand{\rcn}{\cnreduces}
\newcommand{\rcv}{\cvreduces}
\newcommand{\rcnv}{\cnvreduces}
\newcommand{\rcp}{\cpreduces}
\newcommand{\sreduces}{{~\Downarrow^{\stable}}~}
\newcommand{\snreduces}{{~\Downarrow^{\stable}_{\emptyset}}~}
\newcommand{\svreduces}{{~\Downarrow^{\stable}_{\mathrm{ok}}}~}
\newcommand{\snvreduces}{{~\Downarrow^{\stable}_{\emptyset,\mathrm{ok}}}~}
\newcommand{\spreduces}{{~\Downarrow^{\stable}_{\mathrm{det}}}~}
\newcommand{\rs}{\sreduces}
\newcommand{\rsn}{\snreduces}
\newcommand{\rsv}{\svreduces}
\newcommand{\rsnv}{\snvreduces}
\newcommand{\rsp}{\spreduces}

\newcommand{\preduces}{\curvearrowright}
\newcommand{\pc}{~\stackrel{\changeable}{\textstyle \preduces}~}
\newcommand{\ps}{~\stackrel{\stable}{\textstyle \preduces}~}

\newcommand{\ors}{~\downarrow^{\stable}~}
\newcommand{\ods}{~\uparrow^{\stable}~}
\newcommand{\orc}{~\downarrow^{\changeable}~}
\newcommand{\odc}{~\uparrow^{\changeable}~}

\title{A Consistent Semantics of Self-Adjusting Computation}

\author{
Umut A. Acar\footnote{Max-Planck Institute for Software Systems}
\and
Matthias Blume\footnote{Google Inc.}
\and 
Jacob Donham\footnote{Twitter Inc.}
}

\date{June 2011}

\begin{document}
\maketitle

\begin{abstract}
  This paper presents a semantics of self-adjusting computation and
  proves that the semantics are correct and consistent.  The semantics
  integrate change propagation with the classic idea of memoization
  to enable reuse of computations under mutation to memory.  During
  evaluation, reuse of a computation via memoization triggers a change
  propagation that adjusts the reused computation to reflect the
  mutated memory.  Since the semantics integrate memoization and
  change-propagation, it involves both non-determinism (due to
  memoization) and mutation (due to change propagation).  Our
  consistency theorem states that the non-determinism is not harmful:
  any two evaluations of the same program starting at the same state
  yield the same result.  Our correctness theorem states that mutation
  is not harmful: self-adjusting programs are consistent with purely
  functional programming.  We formalize the semantics and their
  meta-theory in the LF logical framework and machine check our proofs
  using Twelf.
\end{abstract}

\section{Introduction}
Many applications operate on data that changes over time.
Self-adjusting computation is a technique that enables program to
respond to changes to their data (e.g., inputs/arguments, external
state, or outcome of tests).  Advances on self-adjusting computation
show that it can speed up response times by orders of magnitude over
recomputing from scratch, closely matching best-known
(problem-specific) algorithms both in theory and in practice
(e.g.,~\cite{AcarBlBlHaTa09}).  More recent results show that the
approach can even enable solving challenging open problems that have
resisted traditional algorithmic approaches
(e.g.~\cite{AcarIhMeSu07,AcarCoHuTu10}).

Key to effectiveness of self-adjusting computation is a technique that
integrates change propagation~\cite{AcarBlHa06}, and the classic idea
of memoization~\cite{Michie68}.  Due to an interesting duality between
memoization and change propagation, combining them turns out to be
crucial for efficiency.  This technique was first developed in two
previously published conference papers.  One paper focused on
algorithmic, implementation, and experimental aspects (journal
version~\cite{AcarBlBlHaTa09}).  The other formal on the formal
aspects and the semantics~\cite{AcarBlDo07}; this paper is a full
version of that conference paper, which it extends by providing full,
machine-checked proofs.  After its publication, the approach proposed
in this paper has essentially served as the foundation for many of the
followup work on self-adjusting computation.  It has been implemented
as a Standard ML library~\cite{AcarBlBlHaTa09} and generalized to
support imperative references~\cite{AcarAhBl08}.  These results set
the stage for the development of the CEAL~\cite{HammerAcCh09} and
Delta ML, which provide direct language support for self-adjusting
computation~\cite{Ley-WildFlAc08}.

Integrating change propagation and memoization poses a major challenge
because the techniques are far from being orthogonal: memoization
traditionally requires purely functional programming, whereas change
propagation is destructive and critically relies on mutation.  Here,
we overcome this challenge by presenting a general semantic framework
that integrates them.  We model memoization as a non-deterministic
oracle; this ensures that the semantics apply to many different ways
in which memoization can be realized.  We prove two main theorems
stating that the semantics are {\em consistent} and {\em correct}
(\secref{correctness}).  The consistency theorem states that the
non-determinism (due to memoization) is harmless by showing that any
two evaluations of the same program in the same store yield the same
result.  The correctness theorem states that self-adjusting
computation is consistent with purely functional programming by
showing that evaluation returns the (observationally) same value as a
purely functional evaluation.  Our proofs do not make any assumptions
about typing.  Our results therefore apply in both typed and untyped
settings.

To study the semantics we extend the {\em adaptive functional
  language}~\afl~\cite{AcarBlHa06}, which support change propagation,
with a construct for memoization.  We call this language \aml
(\secref{language}). The dynamic semantics of \aml{} are store-based.
Mutation to the store between successive evaluations models
incremental changes to the input.  The evaluation of an \aml{} program
also allocates store locations and updates existing locations.  A
memoized expression is evaluated by first consulting the {\em
  memo-oracle}, which non-deterministically returns either a {\em
  miss} or a {\em hit}.  In evaluation, a hit returns a trace of the
evaluation of the memoized expression, which is recursively adapted to
mutations by performing a change propagation on the returned trace.
Intuitively, the idea is to re-use computations (represented via
traces) themselves and recursively perform change propagation on
re-used computations to adapt them according to mutations.  This
contrasts with conventional memoization where results of computations
are re-used in a purely functional (mutation free) setting.

The proofs for the correctness and consistency theorems
(\secref{correctness}) are made challenging because the semantics
consist of a complex set of judgments (where change propagation and
ordinary evaluation are mutually recursive), and because the semantics
involve mutation and two kinds of non-determinism: non-determinism in
memory allocation, and non-determinism due to memoization. Due to
mutation, we are required to prove that evaluation preserves certain
well-formedness properties (e.g., absence of cycles and dangling
pointers).  Due to non-deterministic memory allocation, we cannot
compare the results from different evaluations directly.  Instead, we
compare values structurally by comparing the contents of locations.
To address non-determinism due to memoization, we allow evaluation to
recycle existing memory locations.  Based on these techniques, we
first prove that memoization is harmless: for any evaluation there
exists a memoization-free counterpart that yields the same result
without reusing any computations.  Based on structural equality, we
then show that memoization-free evaluations and fully deterministic
evaluations are equivalent.  These proof techniques may be of
independent interest.

To increase confidence in our results, we encoded the syntax and
semantics of \aml and its meta-theory in the LF logical
framework~\cite{HarperHoPl93} and machine-checked the proofs using
Twelf~\cite{PfenningSc99} (\secref{twelf}).  The Twelf formalization
consist of 7800 lines of code.  The Twelf code is fully foundational:
it encodes all background structures required by the proof and proves
all lemmas from first principles.  We include the full Twelf code in
the appendix (\appref{twelf}).  We note that checking the proofs in
Twelf was not a merely an encoding exercise.  In fact, our initial
attempts at producing a paper-and-pencil proof have failed.  The
process of creating and checking the proof mechanically in Twelf
allowed us to come up with the proof, while also helping us simplify
the rule systems and generalize the proof to untyped languages.  We
therefore feel that the use of Twelf was critical to this result.

Since the semantics model memoization as a non-deterministic oracle,
and since it does not specify how the memory should be allocated while
allowing pre-existing locations to be recycled, the dynamic semantics
of \aml do not translate to an algorithm directly.  In \secref{imp},
we describe some implementation strategies for realizing the \aml
semantics.  

% Variants of these strategies have been implemented
% (e.g.,~\cite{AcarBlBlHaTa09}) since the conference version of this
% work, which omitted the proofs.

\section{The Language}
\label{sec:language}

We describe a language, called \aml, that combines the features of an
adaptive functional language (\afl)~\cite{AcarBlHa06} with
memoization.  The syntax of the language extends that of \afl with
\textbf{memo} constructs for memoizing expressions.  The dynamic
semantics integrate change propagation and evaluation to ensure
correct reuse of computations under mutations.  As explained before,
our results do not rely on typing properties of \aml.  We therefore
omit a type system but identify a minimal set of conditions under
which evaluation is consistent.  In addition to the memoizing and
change-propagating dynamic semantics, we give a pure interpretation of
\aml that provides no reuse of computations.

\subsection{Abstract syntax}
\begin{figure}
\fixedCodeFrame{
\small
\[
\begin{array}{lrcl}

\textit{Values} 
& v & \bnfdef & \unit \bnfalt \Numeral{n} \bnfalt x \bnfalt l \bnfalt (v_1,v_2) \bnfalt \inl{v} \bnfalt \inr{v} \bnfalt \\
&   &         & \sfun{f}{x}{e_s} \bnfalt \cfun{f}{x}{e_c}\\[1mm]

\textit{Prim. Op.} 
& o  & \bnfdef & \Not \bnfalt \Plus \bnfalt \Minus \bnfalt \Equal \bnfalt \Less \bnfalt \ldots \\[1mm]

\textit{Exp.} 
& e  & \bnfdef & e_s \bnfalt e_c \\[1mm]

\textit{St. Exp.} 
& e_s    & \bnfdef  & v \bnfalt \op{v_1,\ldots,v_n} \bnfalt \mod{e_c} \bnfalt \smemo{e_s} \bnfalt \sapply{v_1}{v_2} \bnfalt \\
&        &          & \letin{x}{e_s}{e_s'} \bnfalt \letpin{x_1}{x_2}{v}{e_s} \bnfalt \\
&        &          & \caseofflat{v}{x_1}{e_s}{x_2}{e_s'}\\[1mm]

\textit{Ch. Exp.} 
& e_c    & \bnfdef &  \effect{v} \bnfalt \readin{v}{x}{e_c} \bnfalt \cmemo{e_c} \bnfalt \capply{v_1}{v_2} \bnfalt \\
&        &         &  \letin{x}{e_s}{e_c} \bnfalt \letpin{x_1}{x_2}{v}{e_c} \bnfalt \\
&        &         &  \caseofflat{v}{x_1}{e_c}{x_2}{e_c'}\\[1mm]

\textit{Program} 
& p    & \bnfdef &  e_s
\end{array}
\]
}

\caption{The abstract syntax of \aml.}
\label{fig:language::syntax}
\end{figure}

The abstract syntax of \aml is given in~\figref{language::syntax}.  We
use meta-variables $x$, $y$, and $z$ (and variants) to range over
an unspecified set of variables, and meta-variable $l$ (and
variants) to range over a separate, unspecified set of locations---the
locations are modifiable references.  The syntax of \aml is restricted
to ``$2/3$-cps'', or ``named form'', to streamline the presentation of
the dynamic semantics.

Expressions are classified into three categories: values,
\emph{stable} expressions, and \emph{changeable} expressions.  Values
are constants, variables, locations, and the introduction forms for
sums, products, and functions.  The value of a stable expression is
not sensitive to modifications to the inputs, whereas the value of a
changeable expression may directly or indirectly be affected by
them.

The familiar mechanisms of functional programming are embedded in \aml
as stable expressions.  Stable expressions include the \textcd{let}
construct, the elimination forms for products and sums,
stable-function applications, and the creation of new modifiables.  A
{\em stable function} is a function whose body is a stable expression.
The application of a stable function is a stable expression.  The
expression $\mod{e_c}$ allocates a modifiable reference and
initializes it by executing the changeable expression $e_c$.  Note
that the modifiable itself is stable, even though its contents is
subject to change.  A memoized stable expression is written
$\smemo{e_s}$.

Changeable expressions always execute in the context of an enclosing
$\Mod$-expression that provides the implicit target location that
every changeable expression writes to.  The changeable expression
$\effect{v}$ writes the value $v$ into the target.  The expression
$\readin{v}{x}{e_c}$ binds the contents of the modifiable $v$ to the
variable $x$, then continues evaluation of $e_c$.  A \textcd{read} is
considered changeable because the contents of the modifiable on which
it depends is subject to change.  A {\em changeable function} is a
function whose body is a changeable expression. A changeable function
is stable as a value. The application of a changeable function is a
changeable expression.  A memoized changeable expression is written
$\cmemo{e_c}$.  The changeable expressions include the \textcd{let}
expression for ordering evaluation and the elimination forms for sums
and products.  These differ from their stable counterparts because
their bodies consists of changeable expressions.

%%%%%%%%%%%%%%%%%%%%%%%%%%%%%%%%%%%%%%%%%%%%%%%%%%%%%%%%%%%%%%%%%%%%%%
%% Stores, well-formedness, and lifts
%%%%%%%%%%%%%%%%%%%%%%%%%%%%%%%%%%%%%%%%%%%%%%%%%%%%%%%%%%%%%%%%%%%%%%
\subsection{Stores, well-formed expressions, and lifting}
\label{sec:wf}

\begin{figure}
\small
\[
\begin{array}{|c|}
\hline

\infer{\wfl{v}{\s}{v}{\emptyset}}
{v \in \{\unit,\Numeral{n},x\}}

\quad

\infer{\wfl{l}{\s}{v}{\{l\}\cup L}}
{l\in\dom{\s} ~~~~ \wfl{\s(l)}{\s}{v}{L}}

\quad

\infer{\wfl{(v_1,v_2)}{\s}{(v_1',v_2')}{L_1 \cup L_2}}
{\wfl{v_1}{\s}{v_1'}{L_1} ~~~~ \wfl{v_2}{\s}{v_2'}{L_2}}

\\[2mm]

\infer{\wfl{\mod{e_c}}{\s}{\mod{e_c'}}{L}}
{\wfl{e_c}{\s}{e_c'}{L}}

\quad

\infer{\wfl{\inlr{v}}{\s}{\inlr{v'}}{L}}
{\wfl{v}{\s}{v'}{L}}

\quad

\infer{\wfl{\effect{v}}{\s}{\effect{v'}}{L}}
{\wfl{v}{\s}{v'}{L}}

\\[2mm]

\infer{\wfl{\scfun{f}{x}{e}}{\s}{\scfun{f}{x}{e'}}{L}}
{\wfl{e}{\s}{e'}{L}}

\\[2mm]

\infer{\wfl{\op{v_1,\ldots,v_n}}{\s}{\op{v_1',\ldots,v_n'}}{L_1 \cup \cdots \cup L_n}}
{\wfl{v_1}{\s}{v_1'}{L_1} ~~\cdots~~ \wfl{v_n}{\s}{v_n'}{L_n}}

\\[2mm]

\infer{\wfl{\scapply{v_1}{v_2}}{\s}{\scapply{v_1'}{v_2'}}{L_1 \cup L_2}}
{\wfl{v_1}{\s}{v_1'}{L_1} ~~~~ \wfl{v_2}{\s}{v_2'}{L_2}}

\\[2mm]

\infer{\wfl{\letin{x}{e_1}{e_2}}{\s}{\letin{x}{e_1'}{e_2'}}{L \cup L'}}
{\wfl{e_1}{\s}{e_1'}{L} ~~~~ \wfl{e_2}{\s}{e_2'}{L'}}

\\[2mm]

\infer{\wfl{\letpin{x_1}{x_2}{v}{e}}{\s}{\letpin{x_1}{x_2}{v'}{e'}}{L \cup L'}}
{\wfl{v}{\s}{v'}{L} ~~~ \wfl{e}{\s}{e'}{L'}}

\\[2mm]

\infer{\begin{array}{l}
        (\caseofflat{v}{x_1}{e_1}{x_2}{e_2}),\s \wfarrow \\
        \qquad (\caseofflat{v'}{x_1}{e_1'}{x_2}{e_2'}),L \cup L_1 \cup L_2
      \end{array}
    }
{\wfl{v}{\s}{v'}{L} ~~~~ \wfl{e_1}{\s}{e_1'}{L_1} ~~~~ \wfl{e_2}{\s}{e_2'}{L_2}}

\\[4mm]

\infer{\wfl{\scmemo{e}}{\s}{\scmemo{e'}}{L}}
{\wfl{e}{\s}{e'}{L}}

\\[2mm]

\infer{\wfl{\readin{v}{x}{e_c}}{\s}{\readin{v'}{x}{e_c'}}{L \cup L'}}
{\wfl{v}{\s}{v'}{L} ~~~~ \wfl{e_c}{\s}{e_c'}{L'}}

\\

\hline
\end{array}
\]
\caption{Well-formed expressions and lifts.}
\label{fig:language::wf-lift-full}
\end{figure}

Evaluation of an \aml expression takes place in the context of a
store, written $\s$ (and variants), defined as a finite map from
locations $l$ to values $v$.  We write $\dom{\s}$ for the domain of a
store, and $\s(l)$ for the value at location $l$, provided $l \in
\dom{\s}$.  We write $\s[l \assign v]$ to denote the extension of $\s$
with a mapping of $l$ to $v$.  If $l$ is already in the domain of
$\s$, then the extension replaces the previous mapping.

\[
  \begin{array}{rcl}
    \s[l \assign v](l') &=& \left\{ \begin{array}{l@{~~~~}l}
        v & \mbox{if}~l=l' \\
        \s(l') & \mbox{if}~l\neq l' ~\mbox{and}~ l' \in \dom{\s}
      \end{array}
    \right. \\[2mm]
    \dom{\s[l \assign v]} &=& \dom{\s} \cup \{ l \}
  \end{array}
\]

We say that an expression $e$ is {\em well-formed} in store $\s$ if 1)
all locations reachable from $e$ in $\s$ are in $\dom{\s}$ (``no
dangling pointers''), and 2) the portion of $\s$ reachable from $e$ is
free of cycles.  If $e$ is well-formed in $\s$, then we can obtain a
``lifted'' expression $e'$ by recursively replacing every reachable
location $l$ with its stored value $\s(l)$.  The notion of lifting
will be useful in the formal statement of our main theorems
(\secref{correctness}).

We use the judgment $\wfl{e}{\s}{e'}{L}$ to say that $e$ is
well-formed in $\s$, that $e'$ is $e$ lifted in $\s$, and that $L$ is
the set of locations reachable from $e$ in $\s$.  The rules for
deriving such judgments are shown in \figref{language::wf-lift-full}.
Any finite derivation of such a judgment implies well-formedness of
$e$ in $\s$.

We will use two notational shorthands for the rest of the paper: by
writing $\lift{e}{\s}$ or $\reach{e}{\s}$ we implicitly assert that
there exist a location-free expression $e'$ and a set of locations $L$
such that $\wfl{e}{\s}{e'}{L}$.  The notation $\lift{e}{\s}$ itself
stands for the lifted expression $e'$, and $\reach{e}{\s}$ stands for
the set of reachable locations $L$.  It is easy to see that $e$ and
$\s$ uniquely determine $\lift{e}{\s}$ and $\reach{e}{\s}$ (if they
exist).

%%%%%%%%%%%%%%%%%%%%%%%%%%%%%%%%%%%%%%%%%%%%%%%%%%%%%%%%%%%%%%%%%%%%%%
%% Dynamic Semantics
%%%%%%%%%%%%%%%%%%%%%%%%%%%%%%%%%%%%%%%%%%%%%%%%%%%%%%%%%%%%%%%%%%%%%%
\subsection{Dynamic semantics}
\label{sec:language::dynamic-semantics}

The evaluation judgments of \aml
(\figreftwo{language::dynamic-semantics-stable}{language::dynamic-semantics-changeable})
consist of separate judgments for stable and changeable expressions.
The judgment $\s, e \rs v, \s', \trs$ states that evaluation of the
stable expression $e$ relative to the input store $\s$ yields the
value $v$, the trace $\trs$, and the updated store $\s'$.  Similarly,
the judgment $\s, l \assign e \creduces \s', \trc$ states that
evaluation of the changeable expression $e$ relative to the input
store $\s$ writes its value to the target $l$, and yields the trace
$\trc$ together with the updated store~$\s'$.

A \emph{trace} records the adaptive aspects of evaluation.  Like the
expressions whose evaluations they describe, traces come in stable and
changeable varieties.  The abstract syntax of traces is given by the
following grammar:
\[
\renewcommand{\arraystretch}{1.2}
\begin{array}{lr@{~}c@{~~}l}
\textit{Stable} & \trs &\bnfdef & \epsilon \bnfalt
                                  \trmod{l}{\trc} \bnfalt
                                  \trlet{\trs}{\trs} \\
\textit{Changeable} & \trc &\bnfdef & \treffect{v} \bnfalt
                                      \trlet{\trs}{\trc} \bnfalt
                                      \trread{l}{x}{v}{e}{\trc} \\
\end{array}
\]
A stable trace records the sequence of allocations of modifiables that
arise during the evaluation of a stable expression.  The trace
$\trmod{l}{\trc}$ records the allocation of the modifiable $l$ and the
trace of the initialization code for $l$.  The trace
$\trlet{\trs}{\trs'}$ results from evaluating a \textcd{let}
expression in stable mode, the first trace resulting from the bound
expression, the second from its body.

A changeable trace has one of three forms.  A write, $\treffect{v}$,
records the storage of the value $v$ in the target.  A sequence
$\trlet{\trs}{\trc}$ records the evaluation of a \textcd{let}
expression in changeable mode, with $\trs$ corresponding to the bound
stable expression, and $\trc$ corresponding to its body.  A read
$\trread{l}{x}{v}{e}{\trc}$ specifies the location read ($l$), the
value read ($v$), the context of use of its value ($x.e$) and the
trace ($\trc$) of the remainder of the evaluation within the scope of that
read.  This records the dependency of the target on the value of the
location read.

We define the set of allocated locations of a trace $\tr$, denoted $\gen{\tr}$, as follows:
\[
\begin{array}{|lcl|}
\hline
\gen{\epsilon}              & = & \emptyset\\
\gen{\treffect{v}}          & = & \emptyset\\
\gen{\trmod{l}{\trc}}       & = & \{l\} \cup \gen{\trc}\\
\gen{\trlet{\tri}{\trii}} & = & \gen{\tri} \cup \gen{\trii}\\
\gen{\trread{l}{x}{v}{e}{\trc}}       & = & \gen{\trc}\\
\hline
\end{array} 
\]
For
example, if $\tr_{\mathrm{sample}}~~=~~\Let$
$(\trmod{l_1}{\treffect{2}})~(\trread{l_1}{x}{2}{e}{\treffect{3}})$,
then $\alloc{\tr_{\mathrm{sample}}} = \{ l_1 \}$.

\smallskip
\paragraph{Well-formedness, lifts, and primitive operations.}
We require that primitive operations preserve well-formedness.  In
other words, when a primitive operation is applied to some arguments,
it does not create dangling pointers or cycles in the store, nor does
it extend the set of locations reachable from the argument.  Formally,
this property can be states as follows. 
\[\small
  \begin{array}{l}
    \mbox{If}~\forall i . \wfl{v_i}{\s}{v_i'}{L_i} ~\mbox{and}~
       v=\op{v_1,\ldots,v_n},\\
    \mbox{then}~
      \wfl{v}{\s}{v'}{L}~\mbox{such that}~L \subseteq \bigcup_{i=1}^{n} L_i.
  \end{array}
\]

Moreover, no \aml{} operation is permitted to be sensitive to the
identity of locations.  In the case of primitive operations we
formalize this by postulating that they commute with lifts:
\[\small
  \begin{array}{l}
    \mbox{If}~\forall i . \wfl{v_i}{\s}{v_i'}{L_i} ~\mbox{and}~
       v=\op{v_1,\ldots,v_n},\\
    \mbox{then}~
      \wfl{v}{\s}{v'}{L}~\mbox{such that}~v'=\op{v_1',\ldots,v_n'}.
  \end{array}
\]
In short this can be stated as
$\op{\lift{v_1}{\s},\ldots,\lift{v_n}{\s}} =
\lift{(\op{v_1,\ldots,v_n})}{\s}$.

For example, all primitive operations that operate only on
non-location values preserve well formedness and commute with lifts.

\begin{figure}
\centering
\[
\begin{array}{|c|}
\hline
\infer[\rlabel{valid/s}]
{\s,e_s \rsv v,\s',\tr}
{\begin{array}{c}
    \s,e_s \rs v,\s',\tr \\
    \alloc{\tr} \cap \reach{e_s}{\s} = \emptyset
  \end{array}
}

\qquad

\infer[\rlabel{valid/c}]
{\s,l \assign e_c \rcv \s',\tr}
{\begin{array}{c}
    \s,l \assign e_c \rc \s',\tr \\
    \alloc{\tr} \cap \reach{e_c}{\s} = \emptyset\\
    l \not\in \reach{e_c}{\s} \cup \alloc{\tr}
  \end{array}
}
\\[1mm]
\hline
\end{array}
\]
\caption{Valid evaluations.}
\label{fig:language::well-formed-evaluation}
\label{fig:language::wf-evaluation}
\label{fig:language::wf-eval}
\label{fig:language::valid-eval}
\end{figure}

\smallskip
\paragraph{Valid evaluations.}
We consider only evaluations of well-formed expressions $e$ in stores
$\s$, i.e., those $e$ and $\s$ where $\lift{e}{\s}$ and
$\reach{e}{\s}$ are defined.
Well-formedness is critical for proving correctness: the requirement
that the reachable portion of the store is acyclic ensures that the
approach is consistent with purely functional programming, the
requirement that all reachable locations are in the store ensures that
evaluations do not cause disaster by allocating a ``fresh'' location
that happens to be reachable.  We note that it is possible to omit the
well-formedness requirement by giving a type system and a type safety
proof.  This approach limits the applicability of the theorem only to
type-safe programs.  Because of the imperative nature of the dynamic
semantics, a type safety proof for \aml is also complicated.  We
therefore choose to formalize well-formedness separately.  

Our approach requires showing that evaluation preserves
well-formedness.  To establish well-formedness inductively, we define
{\em valid evaluations}. We say that an evaluation of an expression $e$
in the context of a store $\s$ is {\em valid}, if
\begin{enumerate}
\item $e$ is well-formed in $\s$,
\item the locations allocated during evaluation are disjoint
  from locations that are initially reachable from $e$ (i.e., those
  that are in $\reach{e}{\s}$), and
\item the target location of a changeable evaluation is contained
  neither in $\reach{e}{\s}$ nor the locations allocated during
  evaluation.
\end{enumerate}

\noindent We use $\rsv$ instead of $\rs$ and $\rcv$ instead of $\rc$
to indicate valid stable and changeable evaluations, respectively. The
rules for deriving valid evaluation judgments are shown in
\figref{language::wf-eval}.

\smallskip
\paragraph{The Oracle.}
The dynamic semantics for \aml use an oracle to model memoization.
\figref{language::oracle} shows the evaluation rules for the oracle.
For a stable or a changeable expression $e$, we write an oracle miss
as $\s,e \uparrow^{\stable}$ or $\s, l \assign e_c \uparrow^{\changeable}$, respectively.  The
treatment of oracle hits depend on whether the expression is stable or
changeable.  For a stable expression, it returns the value and the
trace of a valid evaluation of the expression in some store.  For a
changeable expression, the oracle returns a trace of a valid
evaluation of the expression in some store with some destination.

\begin{figure}
\centering
\begin{small}
\[
\begin{array}{|c|}
\hline

\infer[\rlabel{miss/s}]
{\s, e_s \uparrow^{\stable}}
{\strut}

\qquad
\infer[\rlabel{hit/s}]
{\s, e_s \ors v, \tr}
{\s_0, e_s \rsv v, \s_0', \tr}

\\[4mm]

\infer[\rlabel{miss/c}]
{\s, e_c \odc}
{\strut}

\qquad

\infer[\rlabel{hit/c}]
{\s,  e_c \orc \tr}
{\s_0, l \assign e_c \rcv \s_0',\tr}
\\
\hline
\end{array}
\]
\end{small}
\caption{The oracle.}
\label{fig:language::oracle}
\end{figure}

The key difference between the oracle and conventional approaches to
memoization is that the oracle is free to return the trace (and the
value, for stable expressions) of a computation that is consistent
with any store---not necessarily with the current store.  Since the
evaluation whose results are being returned by the oracle can take
place in a different store than the current store, the trace and the
value (if any) returned by the oracle cannot be incorporated into the
evaluation directly.  Instead, the dynamic semantics perform a change
propagation on the trace returned by the oracle before incorporating
it into the current evaluation (this is described below).

\begin{figure*}
\small
\centering
\fixedCodeFrame{
\[
\begin{array}{c}

\infer[\rlabel{value}]
{
\s, v \rs v, \s, \et
}
{
\strut
}

\qquad
\infer[\rlabel{prim.'s}]
{
\s,\op{v_1,\ldots,v_n} \rs v,\s,\et
}             
{
{v = \OpApply{\oper}{(v_1,\ldots,v_n)}}
}
\\[2mm]

%Mod
\infer[\rlabel{mod}]
{
\s,\mod{e} \rs l,\s',\trmod{l}{\tr}
}
{
l \not \in \gen{\tr} \qquad
\s, l \assign e \rc \s', \tr
}
\\[2mm]

\infer[\rlabel{memo/miss}]
{
\s, \smemo{e} \rs v, \s', \tr
}
{
\begin{array}{c}
\s, e \ods\\
\s, e \rs v, \s', \tr
\end{array}
}

\quad

%{Memo}
\infer[\rlabel{memo/hit}]
{
\s, \smemo{e} \rs v, \s', \tr'
}
{
\begin{array}{c}
\s, e \ors v, \tr \\
\s, \tr \ps \s', \tr
'\\
\end{array}
}
\\[2mm]

%{Apply}
\infer[\rlabel{apply}]
{
\s, \sapply{v_1}{v_2} \rs v, \s', \tr
}
{
v_1=\sfun{f}{x}{e} \qquad
\s, [v_1/f,v_2/x]~e \rs v, \s', \tr
}
\\[2mm]

%{Let}
\infer[\rlabel{let}]
{
\s, \letin{x}{e_1}{e_2} \rs v_2, \sii, \trlet{\tri}{\trii}
}
{
\s, e_1  \rs  v_1, \si, \tri  &
\si, [v_1/x]~e_2 \rs  {v_2, \sii, \trii} &
\gen{\tri} \cap \gen{\trii} = \emptyset
}
\\[2mm]

%{Letx}
\infer[\rlabel{let$\cross$}]
{
\s, \letpin{x_1}{x_2}{(v_1,v_2)}{e} \rs v, \s', \tr
}
{
\s, [v_1/x_1,v_2/x_2]~e & \rs & v, \s', \tr
}
\\[2mm]

%Case
\infer[\rlabel{case/inl}]
{
\s, \caseofflat{\inl{v}}{x_1}{e_1}{x_2}{e_2} 
\rs 
v', \s', \tr
}
{
\s, [v/x_1]~e_1 \rs v', \s', \tr
}
\\[2mm]

\infer[\rlabel{case/inr}]
{
\s, \caseofflat{\inr{v}}{x_1}{e_1}{x_2}{e_2} 
\rs
v', \s', \tr
}
{
\s, [v/x_2]~e_2 \rs v', \s', \tr
}\\[-2mm]
\end{array}
\]}
\caption{Evaluation of stable expressions.}
\label{fig:language::dynamic-semantics-stable}
\end{figure*}

\begin{figure*}[t]
\small
\centering
\[
\begin{array}{|c|}
\hline

%Write
\infer[\rlabel{write}]
{
\s, l \assign \effect{v} \rc \s[l \la v], \treffect{v}
}
{
\strut
}
\\[2mm]

%Read
\infer[\rlabel{read}]
{
\s, l \assign \readin{l'}{x}{e} \rc \s',\trread{l'}{x}{\s(l')}{e}{\tr}
}
{
\s, l \assign [\s(l')/x]~e \rc \s', \tr
}        
\\[2mm]

%{Memo}
\infer[\rlabel{memo/miss}]
{
\s, l \assign \cmemo{e} \rc \s', \tr
}
{
\begin{array}{c}
\s, e \odc\\
\s, e \rc  \s', \tr
\end{array}
}
\quad
%{Memo}
\infer[\rlabel{memo/hit}]
{
\s, l \assign \cmemo{e} \rc \s', \tr'
}
{
\begin{array}{c}
\s, e \orc \tr \\
\s, l \assign \tr \pc \s', \tr'\\
\end{array}
}
\\[2mm]

%{Apply}
\infer[\rlabel{apply}]
{
\s, l \assign \capply{v_1}{v_2} \rc \s', \tr
}
{
v_1=\cfun{f}{x}{e} \qquad \s, l \assign [v_1/f,v_2/x]~e \rc \s', \tr
}
\\[2mm]

%Let
\infer[\rlabel{let}]
{
\s, l \assign \letin{x}{e_1}{e_2} \rc \s_2, \trlet{\tri}{\trii}
}
{
\s, e_1  \rs  v, \si, \tri  &
\si, l \assign [v/x]~e_2  \rc  \sii,\trii &
\gen{\tri} \cap \gen{\trii} = \emptyset
}
\\[2mm]

%{Letx}
\infer[\rlabel{let$\cross$}]
{
\s, l \assign \letpin{x_1}{x_2}{(v_1,v_2)}{e} \rc \s', \tr
}
{
\s, l \assign [v_1/x_1,v_2/x_2]~e  \rc  \s',\tr
}
\\[2mm]

%Case
\infer[\rlabel{case/inl}]
{
\s, l \assign \caseofflat{\inl{v}}{x_1}{e_1}{x_2}{e_2}
\rc 
\s', \tr
}
{
\s,l \assign [v/x_1]~e_1 \rc \s',\tr
}
\\[2mm]

\infer[\rlabel{case/inr}]
{
\s,\caseofflat{\inr{v}}{x_1}{e_1}{x_2}{e_2} 
\rc 
\s', \tr
}
{
\s, l \assign [v/x_2]~e_2 \rc \s', \tr
}\\[1mm]
\hline
\end{array}
\]
\caption{Evaluation of changeable expressions.}
\label{fig:language::dynamic-semantics-changeable}
\end{figure*}

\smallskip
\paragraph{Stable Evaluation.}
\figref{language::dynamic-semantics-stable} shows the evaluation rules
for stable expressions.  Most rules are standard for a store-passing
semantics except that they also return traces.  The interesting rules
are those for \Let, \Mod, and \Memo.

The \Let{} rule sequences evaluation of its two expressions, performs
binding by substitution, and yields a trace consisting of the
sequential composition of the traces of its sub-expressions.  For the
traces to be well-formed, the rule requires that they allocate
disjoint sets of locations. The \Mod rule allocates a location $l$,
adds it to the store, and evaluates its body (a changeable expression)
with $l$ as the target.  To ensure that $l$ is not allocated multiple
times, the rule requires that $l$ is not allocated in the trace of the
body.  Note that the allocated location does not need to be fresh---it can
already be in the store, i.e., $l \in \dom{\s}$.  Since every
changeable expression ends with a \Effect, it is guaranteed that an
allocated location is written before it can be read.

The \Memo rule consults an oracle to determine if its body should be
evaluated or not.  If the oracle returns a miss, then the body is
evaluated as usual and the value, the store, and the trace obtained
via evaluation is returned.  If the oracle returns a hit, then it
returns a value $v$ and a trace $\tr$.  To adapt the trace to the
current store $\s$, the evaluation performs a change propagation on
$\tr$ in $\s$ and returns the value $v$ returned by the oracle, and the
trace and the store returned by change propagation.  Note that since
change propagation can change the contents of the store, it can also
indirectly change the (lifted) contents of $v$.

\smallskip 
\paragraph{Changeable Evaluation.}
\figref{language::dynamic-semantics-changeable} shows the evaluation
rules for changeable expressions.  Evaluations in changeable mode
perform {\em destination passing}.  The \Let{}, \Memo{}, \Apply{}
rules are similar to the corresponding rules in stable mode except
that the body of each expression is evaluated in changeable mode.  The
\ReadIn{} expression substitutes the value stored in $\s$ at the
location being read $l'$ for the bound variable $x$ in $e$ and
continues evaluation in changeable mode.  A \Read{} is recorded in the
trace, along with the value read, the variable bound, and the body of
the read.  A \Effect{} simply assigns its argument to the target in
the store.  The evaluation of memoized changeable expressions is
similar to that of stable expressions.

\smallskip 
\paragraph{Change propagation.}
\figref{language::change-propagation} shows the rules for change
propagation.  As with evaluation rules, change-propagation rules are
partitioned into stable and changeable, depending on the kind of the
trace being processed.  The stable change-propagation judgment $\s,
\trs \ps \s', \trs'$ states that change propagating into the stable
trace $\trs$ in the context of the store $\s$ yields the store $\s'$
and the stable trace $\trs'$.  The changeable change-propagation
judgment $\s, l \assign \trc \pc \s', \trc'$ states that change
propagation into the changeable trace $\trc$ with target $l$ in the
context of the store $\s$ yields the changeable trace $\trc'$ and the
store $\s'$.  The change propagation rules mimic evaluation by either
skipping over the parts of the trace that remain the same in the given
store or by re-evaluating the \Read{}s that read locations whose values
are different in the given store.  The rules are labeled with the
expression forms they mimic.

\begin{figure}
\small
\[
\begin{array}{|c|}
\hline
~\qquad\qquad\qquad
\infer[\rlabel{empty}]
{
\s, \et \ps \s, \et
}
{\strut}
\\[-4mm]

%Mod
\infer[\rlabel{mod}]
{
\s, \trmod{l}{\tr} \ps \s', \trmod{l}{\tr'}
}      
{
\begin{array}{rcl}
l &\not\in& \alloc{\tr'} \\
\s, l \assign \tr &\pc& \s', \tr'
\end{array}
}
\quad
%Write
\infer[\rlabel{write}]
{
\s, l \assign \treffect{v} \pc \s[l \la v] , \treffect{v}
}
{\strut}
\\[2mm]

%{Let}
\infer[\rlabel{let/s}]
{
\s, \trlet{\tri}{\trii} \ps \s'', \trlet{\tri'}{\trii'}
}
{
\begin{array}{ccc}
\s, \tri & \ps &  \s', \tri' \\
\s', \trii & \ps &  \s'', \trii'\\
\multicolumn{3}{c}{\gen{\tri'} \cap \gen{\trii'} = \emptyset}
\end{array}
}
\quad
%Let
\infer[\rlabel{let/c}]
{
\s, l \assign (\trlet{\tri}{\trii}) \pc \s'', (\trlet{\tri'}{\trii'})
}
{
\begin{array}{ccc}
\s, \tri & \pc &  \s', \tri' \\
\s', l \assign \trii & \pc & \s'', \trii'\\
\multicolumn{3}{c}{\gen{\tri'} \cap \gen{\trii'} = \emptyset}
\end{array}
}
\\[2mm]

%Read
\infer[\rlabel{read/no ch.}]
{
\s, l \assign \trread{l'}{v}{x}{e}{\tr} \pc \s', \trread{l'}{v}{x}{e}{\tr'}
}
{
\s(l') = v  \qquad
\s, l \assign \tr \pc \s', \tr'
}
\\[2mm]

\infer[\rlabel{read/ch.}]
{
\s, l \assign \trread{l'}{x}{v}{e}{\tr} \pc \s', \trread{l'}{x}{\s(l')}{e}{\tr'}
}
{
\s(l') \not= v \qquad \s, l \assign [\s(l')/x]e \creduces \s', \tr'
}\\[1mm]
\hline
\end{array}
\]
\caption{Change propagation judgments.}
\label{fig:language::cp}
\label{fig:language::change-propagation}
\end{figure}

If the trace is empty, change propagation returns an empty trace and
the same store.  The \Mod rule recursively propagates into the trace
$\tr$ for the body to obtain a new trace $\tr'$ and returns a trace
where $\tr$ is substituted by $\tr'$ under the condition that the
target $l$ is not allocated in $\tr'$.  This condition is necessary to
ensure the allocation integrity of the returned trace.  The stable
\Let rule propagates into its two parts $\tri$ and $\trii$ recursively
and returns a trace by combining the resulting traces $\tri'$ and
$\trii'$ provided that the resulting trace ensures allocation
integrity.
The \Effect rule performs the recorded write in the given store by
extending the target with the value recorded in the trace.  This is
necessary to ensure that the result of a re-used changeable
computation is recorded in the new store.  The \Read{} rule depends on
whether the contents of the location $l'$ being read is the same in
the store as the value $v$ recorded in the trace.  If the contents is
the same as in the trace, then change propagation proceeds into the
body $\tr$ of the read and the resulting trace is substituted for
$\tr$.  Otherwise, the body of the \Read{} is evaluated with the
specified target.  Note that this makes evaluation and
change-propagation mutually recursive---evaluation calls
change-propagation in the case of an oracle hit.  The changeable \Let
rule is similar to the stable \Let.

Most change-propagation judgments perform some consistency checks and
otherwise propagate forward.  Only when a \Read{} finds that the
location in question has changed, it re-runs the changeable
computation that is in its body and replaces the corresponding trace.

\smallskip 
\paragraph{Evaluation invariants.}
Valid evaluations of stable and changeable expressions satisfy the
following invariants:
\begin{enumerate}
\item All locations allocated in the trace are also allocated in the
  result store, i.e., if $\s,e \rsv v,\s',\tr$ or $\s,l \assign e \rcv
  \s',\tr$, then $\dom{\s'} = \dom{\s} \cup \alloc{\tr}$.

\item For stable evaluations, any location whose content changes is
  allocated during that evaluation, i.e., if $\s,e \rsv v,\s',\tr$ and
  $\s'(l) \neq \s(l)$, then $l \in \alloc{\tr}$.
\item For changeable evaluations, a location whose content changes is
  either the target or gets allocated during evaluation, i.e, if
  $\s,l' \assign e \rcv \s',\tr$ and $\s'(l) \neq \s(l)$, then $l \in
  \alloc{\tr} \cup \{ l' \}$.
\end{enumerate}

\smallskip 
\paragraph{Memo-free evaluations.}
The oracle rules introduce non-determinism into the dynamic semantics.
~\lemreftwo{memo-free-stable}{memo-free-changeable} in
\secref{correctness} express the fact that this non-determinism is
harmless: change propagation will correctly update all answers
returned by the oracle and make everything look as if the oracle never
produced any answer at all (meaning that only {\bf memo/miss} rules
were used).

We write $\s,e \rsn v,\s',\tr$ or $\s,l \assign e \rcn \s',\tr$ if
there is a derivation for $\s,e \rs v,\s',\tr$ or $\s,l \assign e \rc
\s',\tr$, respectively, that does not use any {\bf memo/hit} rule.  We
call such an evaluation {\em memo-free}.  We use $\rsnv$ in place of
$\rsv$ and $\rcnv$ in place of $\rcv$ to indicate that a valid
evaluation is also memo-free.

\begin{figure*}
\small
\centering
\[
\begin{array}{|c|}
\hline
%Value & 
\infer[\rlabel{value}]
{ v \rsp v }
{ v & \neq & l }

\qquad

%{Op's}
\infer[\rlabel{prim.}]
{ \op{v_1,\ldots,v_n} \rsp v }
{ v = \OpApply{\oper}{(v_1,\ldots,v_n)} }

\qquad

%Mod
\infer[\rlabel{mod}]
{ \mod{e} \rsp v }
{ e \rcp v }
\\[3mm]

%{Memo}
\infer[\rlabel{memo}]
{ \smemo{e} \rsp v }
{ e \rsp v }

\qquad

%{Apply}
\infer[\rlabel{apply}]
{ \sapply{v_1}{v_2} \rsp v }
{
\begin{array}{c}
(v_1=\sfun{f}{x}{e}) \\
{[v_1/f,v_2/x]~e} \rsp v
\end{array}
}
\\[3mm]

%{Let}
\infer[\rlabel{let}]
{ \letin{x}{e_1}{e_2} \rsp v_2 }
{
\begin{array}{rcl}
e_1 & \rsp &  v_1 \\[1mm]
{[v_1/x]~e_2} & \rsp &  v_2
\end{array}
}
\qquad
%{Letx}
\infer[\rlabel{let$\cross$}]
{ \letpin{x_1}{x_2}{(v_1,v_2)}{e} \rsp v }
{ {[v_1/x_1,v_2/x_2]~e}  \rsp  v }
\\[2mm]

%Case
\infer[\rlabel{case/inl}]
{ \caseof{\inl{v}}{x_1}{e_1}{x_2}{e_2} \rsp v' }
{ {[v/x_1]~e_1} \rsp v' }
\\[4mm]

\infer[\rlabel{case/inr}]
{ \caseof{\inr{v}}{x_1}{e_1}{x_2}{e_2} \rsp v' }
{ {[v/x_2]~e_2} \rsp v' }\\[2mm]
\hline
\end{array}
\]
\[
\begin{array}{|c|}
\hline
%Write
\infer[\rlabel{write}]
{ \effect{v} \rcp v }
{ \strut }

\qquad

%Read
\infer[\rlabel{read}]
{ \readin{v}{x}{e} \rcp v' }
{ {[v/x]~e} \rcp v' }
\\[2mm]

%{Memo}
\infer[\rlabel{memo}]
{ \cmemo{e} \rcp v }
{ e \rcp v }

\quad

%{Apply}
\infer[\rlabel{apply}]
{ \capply{v_1}{v_2} \rcp v }
{\begin{array}{c}
v_1=\cfun{f}{x}{e}\\
{[v_1/f,v_2/x]~e} \rcp v
\end{array}
}
\\[2mm]

%Let
\infer[\rlabel{let}]
{ \letin{x}{e_1}{e_2} \rcp v_2 }
{
\begin{array}{rcl}
e_1 & \rsp &  v_1 \\[1mm]
{[v_1/x]~e_2} & \rcp &  v_2
\end{array}
}

\qquad

%{Letx}
\infer[\rlabel{let$\cross$}]
{ \letpin{x_1}{x_2}{(v_1,v_2)}{e}  \rcp  v }
{ {[v_1/x_1,v_2/x_2]~e}   \rcp  v }
\\[2mm]

%Case
\infer[\rlabel{case/inl}]
{ \caseof{\inl{v}}{x_1}{e_1}{x_2}{e_2}  \rcp  v' }
{ {[v/x_1]~e_1}  \rcp  v' }
\\[4mm]

\infer[\rlabel{case/inr}]
{ \caseof{\inr{v}}{x_1}{e_1}{x_2}{e_2}  \rcp  v' }
{ {[v/x_2]~e_2}  \rcp  v' }\\
\hline
\end{array}
\]

\caption{Purely functional semantics of (location-free) expressions}
\label{fig:language::pure-semantics}
\end{figure*}

\subsection{Deterministic, purely functional semantics}
\label{sec:language::pure}
By ignoring memoization and change-propagation, we can give an
alternative, purely functional, semantics for location-free \aml
programs, which we present in \figref{language::pure-semantics}.
This semantics gives a store-free, pure,
deterministic interpretation of \aml that provides for no computation
reuse.  Under this semantics, both stable and changeable expressions
evaluate to values, \Memo, \Mod and \Effect are simply identities, and
\Read acts as another binding construct.  Our correctness result
states that the pure interpretation of \aml yields results that are
the same (up to lifting) as those obtained by \aml's dynamic semantics
(\secref{correctness}).

\section{Consistency and Correctness}
\label{sec:correctness}

% -*- latex -*-

\newcommand{\llempic}[3]{%
\begin{picture}(1.5,.4)
  \put(0,0){\framebox(1.5,0.4){\ensuremath{%
      \begin{array}{l}
        \mbox{If}~~ {#1} \\[.5mm]
        \mbox{then}~~ {#2}
      \end{array}}
    }}
  \put(0,-0.02){\makebox(0,0)[tl]{#3}}
\end{picture}}

\newcommand{\rlempic}[3]{%
\begin{picture}(1.5,.4)
  \put(0,0){\framebox(1.5,0.4){\ensuremath{%
      \begin{array}{l}
        \mbox{If}~~ {#1} \\[.5mm]
        \mbox{then}~~ {#2}
      \end{array}}
    }}
  \put(1.5,-0.02){\makebox(0,0)[tr]{#3}}
\end{picture}}

\begin{figure}[t]
\setlength{\unitlength}{1.25in}
\begin{center}

\begin{picture}(3.25,2)

\put(0,0){\dashbox{0.02}(3.25,2){\begin{picture}(3.25,2)

\put(0.075,0.05){\begin{picture}(3.1,1.9)

\put(-0.05,0.6){\dashbox{0.01}(1.55,1.3){\begin{picture}(1.55,1.3)

%% Lemma memo-elim (1)
\put(0.025,0.8){\llempic{\s,e \rsv v_1,\s_1,\tr_1}
                    {\s,e \rsnv v_1,\s_1,\tr_1}
                    {\lemref{memo-free-stable}}}

%% Lemma purity (1)
\put(0.025,0.2){\llempic{s,e \rsnv v_1,\s_1,\tr_1}
                  {(e\!\uparrow\!\s) \rsp (v_1\!\uparrow\!\s_1)}
                  {\lemref{purity-st}}}

\put(0.775,0.8){\vector(0,-1){.2}}
\end{picture}}}
\put(0,0.58){\makebox(0,0)[tl]{\thmref{purity}}}

\put(1.6,0.6){\dashbox{0.01}(1.55,1.3){\begin{picture}(1.55,1.3)

%% Lemma memo-elim (2)
\put(0.025,0.8){\rlempic{\s,e \rsv v_2,\s_2,\tr_2}
                    {\s,e \rsnv v_2,\s_2,\tr_2}
                    {\lemref{memo-free-stable}}}

%% Lemma purity (2)
\put(0.025,0.2){\rlempic{\s,e \rsnv v_2,\s_2,\tr_2}
                  {(e\!\uparrow\!\s) \rsp (v_2\!\uparrow\!\s_2)}
                  {\lemref{purity-st}}}

\put(0.755,0.8){\vector(0,-1){.2}}
\end{picture}}}
\put(3.1,0.58){\makebox(0,0)[tr]{\thmref{purity}}}

%% Deterministic semantics
\put(0.55,0){\framebox(2,.4){\(
\begin{array}{ll}
\mbox{But since}~ \rsp ~\mbox{is deterministic}, \\
\mbox{it follows that}~ (v_1\!\uparrow\!\s_1) = (v_2\!\uparrow\!\s_2)
\end{array}\)
}}

%% diagonal arrows
\put(0.75,0.6){\vector(1,-1){.2}}
\put(2.4,0.6){\vector(-1,-1){.2}}

\end{picture}}

\end{picture}}}

\put(3.25,-0.02){\makebox(0,0)[rt]{\thmref{consistency}}}

\end{picture}

\end{center}
\caption{The  structure of the proofs.}
\label{fig:correctness::proof-structure}
\end{figure}
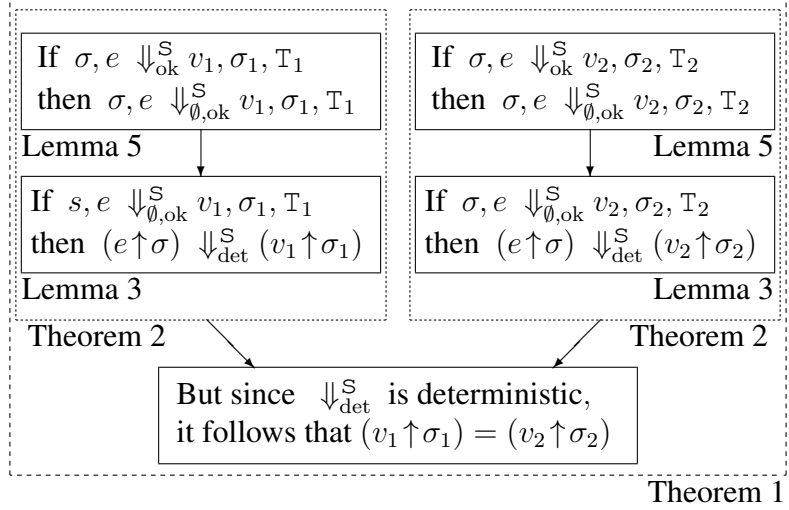

We now state consistency and correctness theorems for \aml and outline
their proofs in terms of several main lemmas.  As depicted in
\figref{correctness::proof-structure}, consistency
(\thmref{consistency}) is a consequence of correctness
(\thmref{purity}).

\subsection{Main theorems}

Consistency uses {\em structural equality} based on the notion of {\em
  lifts} (see \secref{wf}) to compare the results of two potentially
different evaluations of the same \aml program under its
non-deterministic semantics.  Correctness, on the other hand, compares
one such evaluation to a pure, functional evaluation.  It justifies
saying that even with stores, memoization and change propagation,
\aml{} is essentially a purely functional language.

\begin{theorem}[Consistency]
\label{thm:consistency}
If $\s,e \rsv  v_1,\s_1,\tr_1$ and $\s,e \rsv  v_2,\s_2,\tr_2$, then
$\lift{v_1}{\s_1} = \lift{v_2}{\s_2}$.
\end{theorem}

\begin{theorem}[Correctness]
\label{thm:purity}
If $\s,e \rsv v,\s',\tr$, then $(e \uparrow \s) \rsp (v \uparrow \s')$.
\end{theorem}

Recall that by our convention the use of the notation $\lift{v}{\s}$
implies well-formedness of $v$ in $\s$.  Therefore, part of the
statement of consistency is the preservation of well-formedness during
evaluation, and the inability of \aml{} programs to create cyclic
memory graphs.

\subsection{Proof outline}

The consistency theorem is proved in two steps.  First,
\lemreftwo{purity-st}{purity-ch} state that consistency is true in the
restricted setting where all evaluations are memo-free.

\begin{lemma}[purity/st.]
\label{lem:purity-st}
If $\s,e \rsnv v,\s',\tr$, then $(e \uparrow \s) \rsp (v \uparrow \s')$.
\end{lemma}

\begin{lemma}[purity/ch.]
\label{lem:purity-ch}
If $\s,l \assign e \rcnv \s',\tr$, then $(e \uparrow \s) \rcp (l \uparrow \s')$.
\end{lemma}

Second, \lemreftwo{memo-free-stable}{memo-free-changeable} state
that for any evaluation there is a memo-free counterpart that yields
an {\em identical} result and has {\em identical} effects on the store.
Notice that this is stronger than saying that the memo-free evaluation
is ``equivalent'' in some sense (e.g., under lifts).
The statements of these lemmas are actually even stronger since they
include a ``preservation of well-formedness'' statement.  Preservation
of well-formedness is required in the inductive proof.

\begin{lemma}[memo-freedom/st.]
\label{lem:memo-free-stable}
If $\s,e \rsv v,\s',\tr$,
then $\s,e \rsn v,\s',\tr$ where $\reach{v}{\s'} \subseteq \reach{e}{\s}
\cup \alloc{\tr}$.
\end{lemma}

\begin{lemma}[memo-freedom/ch.]
\label{lem:memo-free-changeable}
If $\s,l \la e \rcv \s',\tr$,
then $\s,l \la e \rcn \s',\tr$ where
$\reach{\s'(l)}{\s'} \subseteq \reach{e}{\s} \cup \alloc{\tr}$.
\end{lemma}

The proof for~\lemreftwo{memo-free-stable}{memo-free-changeable}
proceeds by simultaneous induction over the expression $e$.  It is
outlined in far more detail in~\secref{proof}.
Both lemmas state that if there is a well-formed
evaluation leading to a store, a trace, and a result (the value $v$ in
the stable lemma, or the target $l$ in the changeable lemma), the same
result (which will be well-formed itself) is obtainable by a memo-free
run.  Moreover, all locations reachable from the result were either
reachable from the initial expression or were allocated during the
evaluation.  These conditions help to re-establish well-formedness in
inductive steps.

The lemmas are true thanks to a key property of the dynamic semantics:
allocated locations need not be completely ``fresh'' in the sense that
they may be in the current store as long as they are neither reachable
from the initial expression nor get allocated multiple times.  This
means that a location that is already in the store can be chosen for
reuse by the \Mod expression (\figref{language::dynamic-semantics-stable}).
To see why this is important, consider as an example the evaluating of the
expression: $\smemo{(\mod(\effect{3}))}$ in $\s$.  Suppose now that
the oracle returns the value $l$ and the trace $T_0$:
\( \s_0,\mod(\effect{3}) \rs l,\s_0',\tr_0.
\)
Even if $l \in \dom{\s}$,
change propagation will simply update the store as $\s[l \la 3]$ and
return $l$.  In a memo-free evaluation of the same expression
the oracle misses, and \Mod must allocate
a location.  Thus, if the evaluation of \Mod were restricted to use
fresh locations only, it would allocate some $l' \not\in \dom{\s}$,
and return that.  But since $l \in \dom{\s}$, $l \neq l'$.

\section{The Proofs}
\label{sec:proof}
\label{app:proof}
This sections presents a proof sketch for the four memo-elimination
lemmas as well as the two lemmas comparing \aml's dynamic semantics to
the pure semantics (\secref{correctness}).  We give a detailed
analysis for the most difficult cases.  These proofs have all been
formalized and machine-checked in Twelf (see \secref{twelf}).

\subsection{Proofs for memo-elimination}
\label{sec:proof::memo-elim}

Informally speaking, the proofs
for~\lemreftwo{memo-free-stable}{memo-free-changeable}, as well as
\lemreftwo{hit-elim-stable}{hit-elim-changeable} all proceed by
simultaneous induction on the derivations of the respective {\em
  result} evaluation judgments.  The imprecision in this statement
stems from the fact that, as we will see, there are instances where we
use the induction hypothesis on something that is not really a
sub-derivation of the given derivation.  For this reason, a full
formalization of the proof defines a metric on derivations which
demonstrably decreases on each inductive step.  The discussion of the
formalization in Twelf in \secref{twelf} has more details
on this.

\subsubsection*{Substitution}

We will frequently appeal to the following {\em substitution lemma}.
It states that well-formedness and lifts of expressions are preserved
under substitution:

\begin{lemma}[Substitution]
\label{lem:substitution}
If $\wfl{e}{\s}{e'}{L}$ and $\wfl{v}{\s}{v'}{L'}$, then
$\wfl{[v/x]~e}{\s}{[v'/x]~e'}{L''}$ with $L'' \subseteq L \cup L'$.
\end{lemma}

\noindent The proof for this proceeds by induction on the structure of $e$.

\subsubsection*{Hit-elimination lemmas}

Since the cases for the {\bf memo/hit} rules involve many sub-cases,
it is instructive to separate these out into separate lemmas:

\begin{lemma}[hit-elimination/stable]
\label{lem:hit-elim-stable}
If $\s_0,e \rsv v,\s_0',\tr_0$ and $\s,\tr_0 \ps \s',\tr$ where
$\reach{e}{\s} \cap \alloc{\tr} = \emptyset$, \\
then $\s,e \rsn v,\s',\tr$ with $\reach{v}{\s'} \subseteq \reach{e}{\s}
\cup \alloc{\tr}$.
\end{lemma}

\begin{lemma}[hit-elimination/changeable]
\label{lem:hit-elim-changeable}
If $\s_0,l_0 \assign e \rcv \s_0',\tr_0$ and $\s,l \assign \tr_0 \pc
\s',\tr$ where $\reach{e}{\s} \cap \alloc{\tr} = \emptyset$
and
$l \not\in \reach{e}{\s} \cup \alloc{\tr}$, \\
then $\s,l \assign e \rcn \s',\tr$ with $\reach{\s'(l)}{\s'}
\subseteq \reach{e}{\s} \cup \alloc{\tr}$.
\end{lemma}

\subsubsection*{Proof sketch for~\lemref{memo-free-stable} (stable memo-freedom)}
For the remainder of the current section we will ignore the added
complexity caused by the need for a decreasing metric on derivations.
Here is a sketch of the cases that need to be considered in the part
of the proof that deals with \lemref{memo-free-stable}:

\begin{itemize}
\item{\bf value:} Since the expression itself is the value, with the trace
  being empty, this case is trivial.

\item{\bf primitives:} The case for primitive operations goes through
  straightforwardly using preservation of well-formedness.
  
\item{\bf mod:} Given $\s,\mod{e} \rsv l,\s',\trmod{l}{\tr}$ we have
  \[\reach{\mod{e}}{\s} \cap \alloc{\trmod{l}{\tr}} = \emptyset.\]
  This implies that $l \not\in \reach{\mod{e}}{\s}$.  By the
  evaluation rule {\bf mod} it is also true that $\s,e \rc \s',\tr$
  and $l \not\in \alloc{\tr}$.  By definition of $\mathcd{reach}$ and
  $\mathcd{alloc}$ we also know that $\reach{e}{\s} \cap \alloc{\tr} =
  \emptyset$, implying $\s,e \rcv \s',\tr$.
  
  By induction (using~\lemref{memo-free-changeable}) we get
  $\s,l \assign e \rcn \s',\tr$ with $\reach{\s'(l)}{\s'} \subseteq
  \reach{e}{\s} \cup \alloc{\tr}$.  Since $l$ is the final result, we
  find that
  \begin{eqnarray*}
    \reach{l}{\s'} &=& \reach{\s'(l)}{\s'} \cup \{ l \} \\
                   &\subseteq& \reach{e}{\s} \cup \alloc{\tr} \cup \{ l \} \\
                   &=& \reach{e}{\s} \cup \alloc{\trmod{l}{\tr}}.
  \end{eqnarray*}

\item{\bf memo/hit:} Since the result evaluation is supposed to be
  memo-free, there really is no use of the {\bf memo/hit} rule there.
  However, a {\bf memo/miss} in the memo-free trace can be the result
  of eliminating a {\bf memo/hit} in the original run.  We refer to
  this situation here, which really is the heart of the matter: a use
  of the {\bf memo/hit} rule for which we have to show that we can
  eliminate it in favor of some memo-free evaluation.  This case has
  been factored out as a separate lemma
  (\lemref{hit-elim-stable}), which we can use here
  inductively.

\item{\bf memo/miss} The case of a retained {\bf memo/miss} is completely
  straightforward, using the induction hypothesis
  (\lemref{memo-free-stable}) on the subexpression $e$ in
  $\mod{e}$.

\item{\bf let} The difficulty here is to establish that the second part of
  the evaluation is valid. Given
  \[
  \s,\letin{x}{e_1}{e_2} \rsv v_2,\s'',\trlet{\tr_1}{\tr_2}
  \]
  we have
  \(L \cap \alloc{\trlet{\tr_1}{\tr_2}} = \emptyset\) \\
  where
  \(L = \reach{\letin{x}{e_1}{e_2}}{\s}.\)

  By the evaluation rule {\bf let} it is the case that
  \(\s,e_1 \rs v_1,\s',\tr_1\)
  where
  \(\alloc{\tr_1} \subseteq \alloc{\tr}.\)
  Well-formedness of the whole expression implies well-formedness of
  each of its parts, so
  $\reach{e_1}{\s} \subseteq L$ and $\reach{e_2}{\s} \subseteq L$.
  This means that $\reach{e_1}{\s} \cap \alloc{\tr_1} = \emptyset$, so $\s,e_1
  \rsv v_1,\s',\tr_1$.  Using the induction hypothesis
  (\lemref{memo-free-stable}) this implies
  \[\s,e_1 \rsn v_1,\s',\tr_1\]
  and
  \(\reach{v_1}{\s'} \subseteq \reach{e_1}{\s} \cup \alloc{\tr_1}.\)
  
  Since $\reach{e_2}{\s} \subseteq L$ we have $\reach{e_2}{\s} \cap
  \alloc{\tr_1} = \emptyset$.  Store $\s'$ is equal to $\s$ up to
  $\alloc{\tr_1}$, so $\reach{e_2}{\s} = \reach{e_2}{\s'}$.
  Therefore, by substitution (\lemref{substitution}) we
  get
  \begin{eqnarray*}
    \reach{[v_1/x]~e_2}{\s'} &\subseteq&
       \reach{e_2}{\s'} \cup \reach{v_1}{\s'} \\
     &\subseteq&
       \reach{e_2}{\s} \cup \reach{v_1}{\s'} \\
     &\subseteq&
       \reach{e_2}{\s} \cup \reach{e_1}{\s} \\
       & & \cup \alloc{\tr_1} \\
     &=& L \cup \alloc{\tr_1}
  \end{eqnarray*}

  Since $\alloc{\tr_2}$ is disjoint from both $L$ and $\alloc{\tr_1}$,
  this means that
  \(\s',[v_1/x]~e_2 \rsv v_2,\s'',\tr_2\).
  Using the induction hypothesis (\lemref{memo-free-stable})
  a second time we get
  \[\s',[v_1/x]~e_2 \rsn v_2,\s'',\tr_2,\]
  so by definition
  \[\s,\letin{x}{e_1}{e_2} \rsn v_2,\s'',\trlet{\tr_1}{\tr_2}.\]
  It is then also true that
  \begin{eqnarray*}
    \reach{v_2}{\s''} &\subseteq& \reach{[v_1/x]~e_2}{\s'} \cup \alloc{\tr_2} \\
                      &\subseteq& L \cup \alloc{\tr_1} \cup \alloc{\tr_2} \\
                      &=& L \cup \alloc{\trlet{\tr_1}{\tr_2}},
  \end{eqnarray*}
  \noindent which concludes the argument.
\end{itemize}

The remaining cases all follow by a straightforward application of
\lemref{substitution} (substitution), followed by the use of the
induction hypothesis (\lemref{memo-free-stable}).

\subsubsection*{Proof sketch for~\lemref{memo-free-changeable} (Changeable memo-freedom)}

\begin{itemize}
\item{\bf write:} Given $\s,l \assign \effect{v} \rcv \s[l \la
  v],\treffect{v}$ we clearly also have $\s,l \assign \effect{v} \rcn
  \s[l \la v],\treffect{v}$.  First we need to show that $\s'(l)$ is
  well-formed in $s' = \s[l \la v]$.  This is true because $\s'(l) =
  v$ and $l$ is not reachable from $v$ in $\s$, so the update to $l$
  cannot create a cycle.  Moreover, this means that the locations
  reachable from $v$ in $\s'$ are the same as the ones reachable in
  $\s$, i.e., $\reach{v}{\s} = \reach{v}{\s'}$.  Since nothing is
  allocated, $\alloc{\treffect{v}} = \emptyset$, so obviously
  $\reach{\s'(l)}{\s'} \subseteq \reach{v}{\s} \cup \alloc{\treffect{v}}$.

\item{\bf read:} For the case of $\s,l \assign \readin{l'}{x}{e} \rcv
  \s',\tr$ we observe that by definition of well-formedness $\s(l')$ is
  also well-formed in $\s$.  From here the proof proceeds by an
  application of the substitution lemma, followed by a use of the
  induction hypothesis (\lemref{memo-free-changeable}).

\item{\bf memo/hit:} Again, this is the case of a {\bf memo/miss} which is
  the result of eliminating the presence of a {\bf memo/hit} in the
  original evaluation.  Like in the stable setting, we have factored
  this out as a separate lemma
  (\lemref{hit-elim-changeable}).

\item{\bf memo/miss:} As before, the case of a retained use of {\bf
    memo/miss} is handled by straightforward use of the induction
  hypothesis (\lemref{memo-free-changeable}).

\item{\bf let:} The proof for the {\bf let} case in the changeable
  setting is tedious but straightforward and proceeds along the lines
  of the proof for the {\bf let} case in the stable setting.
  \lemref{memo-free-stable} is used inductively for the first
  sub-expression, \lemref{memo-free-changeable} for the second (after
  establishing validity using the substitution lemma).
\end{itemize}

The remaining cases follow by application of
the substitution lemma and the use of the induction
hypothesis (\lemref{memo-free-changeable}).

\subsubsection*{Proof of ~\lemref{hit-elim-stable} (stable hit-elimination)}

\begin{itemize}
\item{\bf value:} Immediate.
\item{\bf primitives:} Immediate.
\item{\bf mod:} The case of {\bf mod} requires some attention, since
  the location being allocated may already be present in $\s$, a situation
  which, however, is tolerated by our relaxed evaluation rule for
  $\mod{e}$.  We show the proof in detail, using the following
  calculations which establishes the conclusions (lines $(16,19)$)
  from the preconditions (lines $(1,2,3)$):

  \[
  \small
  \begin{array}{l@{~}l@{~~}r@{~}c@{~}l}
    (1) &                      & \s_0,\mod{e} &\rsv& l,\s_0',\trmod{l}{\tr_0} \\
    (2) &                      & \s,\trmod{l}{\tr_0} &\ps& \s',\trmod{l}{\tr} \\
    (3) &                      & \multicolumn{3}{c}{\reach{e}{\s} \cap
                                     \alloc{\tr} = \emptyset} \\
        &                      & \multicolumn{3}{c}{l \not\in
                                    \alloc{\tr} \cup \reach{e}{\s}} \\[1mm]
    (4) &\by (1)               & \s_0,l \assign e &\rc& \s_0',\tr_0 \\
    (5) &\by (1)               & \multicolumn{3}{c}{\alloc{\trmod{l}{\tr_0}} \cap \reach{e}{\s_0} = \emptyset} \\
    (6) &\by (5)               & \multicolumn{3}{c}{\alloc{\tr_0} \cap \reach{e}{\s_0} = \emptyset} \\
    (7) &\by (5)               & l &\not\in& \reach{e}{\s_0} \\
    (8) &\by (1),\mathbf{mod}  & l &\not\in& \alloc{\tr_0} \\[1mm]
    (9) &\by (4,6,7,8)         & \s_0,l \assign e &\rcv& \s_0',\tr_0 \\
    (10) &\by (2),\mathbf{\mod}& \s,l \assign \tr_0 &\pc& \s',\tr \\
    (11) &\by (3)              & \multicolumn{3}{c}{\reach{e}{\s} \cap \alloc{\tr} = \emptyset} \\
    (12) &\by (3)              & l &\not\in& \reach{e}{\s} \\
    (13) &\by (3)              & l &\not\in& \alloc{\tr} \\[1mm]
    (14) &\by (9-13),\mbox{IH} & \s,l \assign e &\rcn& \s',\tr \\
    (15) &\by (9-13),\mbox{IH} & \multicolumn{3}{c}{
      \begin{array}{rcl}
        \reach{\s'(l)}{\s'} &\subseteq& \reach{e}{\s} \cup
                                        \alloc{\tr}
    \end{array}} \\[1mm]
    (16) &\by (8,14),\mathbf{mod}& \s,\mod{e} &\rsn& l,\s',\trmod{l}{\tr} \\
    (17) &\by (7,8,15)         & l &\not\in& \reach{\s'(l)}{\s'} \\
    (18) &\by (17)             & \multicolumn{3}{c}{\reach{l}{\s'} = \reach{\s'(l)}{\s'} \cup \{ l \}} \\
    (19) &\by (15,18)          &
      \multicolumn{3}{l}{
        \begin{array}{rcl}
          \reach{l}{\s'} &\subseteq& \reach{e}{\s} \cup 
                                     \alloc{\tr} \cup \{ l \} \\
                         &=& \reach{e}{\s} \cup 
                              \alloc{\trmod{l}{\tr}}
         \end{array}}
  \end{array}
  \]

\item{\bf memo/hit:} This case is proved by two consecutive applications of
  the induction hypothesis, one time to obtain a memo-free version of
  the original evaluation $\s_0,e \rsn v,\s_0',\tr_0$, and then
  starting from that the memo-free final result.
  
  It is here where straightforward induction on the derivation breaks
  down, since the derivation of the memo-free version of the original
  evaluation is not a sub-derivation of the overall derivation.  In
  the formalized and proof-checked version (\secref{twelf}) this is
  handled using an auxiliary metric on derivations.

\item{\bf memo/miss:} The case where the original evaluation of $\smemo{e}$
  did not use the oracle and evaluated $e$ directly, we prove the result
  by applying the induction hypothesis (\lemref{hit-elim-stable}).

\item{\bf let:} We consider the evaluation of $\letin{x}{e_1}{e_2}$.
  Again, the main challenge here is to establish that the evaluation
  of $[v_1/x]~e$, where $v_1$ is the result of $e_1$, is well-formed.
  The argument is tedious but straightforward and proceeds much like
  that in the proof of \lemref{memo-free-stable}.
\end{itemize}

All remaining cases are handled simply by applying the substitution
lemma (\lemref{substitution}) and then using the induction hypothesis
(\lemref{hit-elim-stable}).  

\subsubsection*{Proof of~\lemref{hit-elim-changeable} (changeable hit-elimination)}

\begin{itemize}
\item{\bf write:} We have $e = \effect{v}$ and $\tr_0 = \tr =
  \treffect{v}$.  Therefore, trivially, $\s,l \assign e \rcn \s',\tr$
  with $\s' = \s[l \la v]$.  Also, $\reach{\effect{v}}{\s} =
  \reach{v}{\s} = L$.  Therefore, $\reach{\s'(l)}{\s'} = L$ because $l
  \not\in L$.  Of course, $L \subseteq L \cup \alloc{\tr}$.
\item{\bf read/no ch.:} We handle {\bf read} in two parts.  The first part
  deals with the situation where there is no change to the location
  that has been read.  In this case we apply the substitution lemma to
  establish the preconditions for the induction hypothesis and
  conclude using~\lemref{hit-elim-changeable}.
\item{\bf read/ch.:} If change propagation detects that the location
  being read contains a new value, it re-executes the body of
  $\readin{l'}{x}{e}$.  Using substitution we establish the
  pre-conditions of~\lemref{memo-free-changeable} and conclude by
  using the induction hypothesis.
\item{\bf memo/hit:} Like in the proof for
  \lemref{hit-elim-stable}, the {\bf memo/hit} case is
  handled by two cascading applications of the induction hypothesis
  (\lemref{hit-elim-changeable}).
\item{\bf memo/miss:} Again, the case where the original evaluation did not
  get an answer from the oracle is handled easily by using the
  induction hypothesis (\lemref{hit-elim-changeable}).
\item{\bf let:} We consider the evaluation of $\letin{x}{e_1}{e_2}$.
  As before, the challenge is to establish that the evaluation
  of $[v_1/x]~e$, where $v_1$ is the (stable) result of $e_1$, is well-formed.
  The argument is tedious but straightforward and proceeds much like
  that in the proof of \lemref{memo-free-changeable}.
\end{itemize}

All remaining cases are handled by the induction hypothesis
(\lemref{hit-elim-changeable}) which becomes applicable after
establishing validity using the substitution lemma.

\subsection{Proofs for equivalence to pure semantics}
The proofs for \lemreftwo{purity-st}{purity-ch}
proceed by simultaneous induction on the derivation of the memo-free
evaluation.  The following two subsections outline the two major parts
of the case analysis.

\subsubsection*{Proof sketch for \lemref{purity-st} (stable evaluation)}

We proceed by considering each possible stable evaluation rule:

\begin{itemize}
\item{\bf value:} Immediate.
\item{\bf primitives:} Using the condition on primitive operations that
  they commute with lifts, this is immediate.
\item{\bf mod:} Consider $\mod{e_c}$. The induction hypothesis
  (\lemref{purity-ch}) on the evaluation of $e_c$
  directly gives the required result.
\item{\bf memo:} Since we consider memo-free evaluations, we only need
  to consider the use of the {\bf memo/miss} rule.  The result follows
  by direct application of the induction hypothesis
  (\lemref{purity-st}).
\item{\bf let:} We have $\s,\letin{x}{e_1}{e_2} \rsn
  v_2,\s'',\trlet{\tr_1}{\tr_2}$.  Because of validity of the original
  evaluation, we also have $\wf{\letin{x}{e_1}{e_2}}{\s}{L}$ with $L
  \cap \alloc{\trlet{\tr_1}{\tr_2}} = \emptyset$.  Therefore, $\s,e_1
  \rsn v_1,\s',\tr_1$ where $\wf{e_1}{\s}{L_1}$ and $L_1 \cap
  \alloc{\tr} = \emptyset$ because $L_1 \subseteq L$ and
  $\alloc{\tr_1} \subseteq \alloc{\trlet{\tr_1}{\tr_2}}$.  By
  induction hypothesis (\lemref{purity-st}) we get $(e_1 \uparrow \s)
  \rsp (v_1 \uparrow \s')$.
  
  We can establish validity for $\s',[v_1/x]~e_2 \rsn v_2,\s'',\tr_2$
  the same way we did in the proof of
  \lemref{memo-free-stable}, so by a second application of the
  induction hypothesis we get $([v_1/x]~e_2 \uparrow \s') \rsp (v_2
  \uparrow \s'')$.  But by substitution
  (\lemref{substitution}) we have $([v_1/x]~e_2) \uparrow \s'
  = [(v_1 \uparrow \s') / x]~(e_2 \uparrow \s')$.  Using the
  evaluation rule {\bf let/p} this gives the desired result.
\end{itemize}

The remaining cases follow straightforwardly by applying the induction
hypothesis (\lemref{purity-st}) after establishing validity
using the substitution lemma.

\subsubsection*{Proof sketch for \lemref{purity-ch} (changeable evaluation)}

here we consider each possible changeable evaluation rule:

\begin{itemize}
\item{\bf write:} Immediate by the definition of lift.
\item{\bf read:} Using the definition of lift and the substitution lemma,
  this follows by an application of the induction hypothesis
  (\lemref{purity-ch}).
\item{\bf memo:} Like in the stable setting, this case is handled by
  straightforward application of the induction hypothesis because no
  memo hit needs to be considered.
\item{\bf let:} The let case is again somewhat tedious.  It proceeds
  by first using the induction hypothesis (\lemref{purity-st}) on the
  stable sub-expression, then re-establishing validity using the
  substitution lemma, and finally applying the induction hypothesis a
  second time (this time in form of \lemref{purity-ch}).
\end{itemize}

All other cases are handled by an application of the induction
hypothesis (\lemref{purity-ch}) after establishing validity
using the substitution lemma.

\section{Mechanization in Twelf}
\label{sec:twelf}

To increase our confidence in the proofs for the correctness and the
consistency theorems, we have encoded the \aml language and the proofs
in Twelf~\cite{PfenningSc99} and machine-checked the proofs.  We
follow the standard \emph{judgments as types}
methodology~\cite{HarperHoPl93}, and check our theorems using the
Twelf metatheorem checker. For full details on using Twelf in this way
for proofs about programming languages, see Harper and Licata's
paper~\cite{HarperLi07}.

The LF encoding of the syntax and semantics of \aml corresponds very
closely to the paper judgments (in an informal sense; we have not
proved formally that the LF encoding is {\em adequate}, and take
adequacy to be evident).  However, in a few cases we have altered the
judgments, driven by the needs of the mechanized proof.  For example,
on paper we write memo-free and general evaluations as different
judgments, and silently coerce memo-free to general evaluations in the
proof.  We could represent the two judgments by separate LF type
families, but the proof would then require a lemma to convert one
judgment to the other.  Instead, we define a type family to represent
general evaluations, and a separate type family, indexed by evaluation
derivations, to represent the judgment that an evaluation derivation
is memo-free.

The proof of consistency (a metatheorem in Twelf) corresponds closely
to the paper proof in overall structure.  The proof of memo-freedom
consists of four mutually-inductive lemmas: memo-freedom for stable
and changeable expressions (\lemref{memo-free-stable} and
\lemref{memo-free-changeable}), and versions of these with an
additional change propagation following the evaluation (needed for the
hit cases).  In the hit cases for these latter lemmas, we must
eliminate two change propagations: we call the lemma once to eliminate
the first, then a second time on the output of the first call to
eliminate the second.  Since the evaluation in the second call is not
a subderivation of the input, we must give a separate termination
metric.  The metric is defined on evaluation derivations and simply
counts the number of evaluations in the derivations, including those
inside of change propagations.  In an evaluation which contains change
propagations, there are ``garbage'' evaluations which are removed
during hit-elimination.  Therefore, hit-elimination reduces this
metric (or keeps it the same, if there were no change propagations to
remove). We add arguments to the lemmas to account for the metric, and
simultaneously prove that the metric is smaller in each inductive
call, in order for Twelf to check termination.

Aside from this structural difference due to termination checking, the
main difference from the paper proof is that the Twelf proof must of
course spell out all the details which the paper proof leaves to the
reader to verify.  In particular, we must encode ``background''
structures such as finite sets of locations, and prove relevant
properties of such structures.  While we are not the first to use
these structures in Twelf, Twelf has poor support for reusable
libraries at present.  Moreover, our needs are somewhat specialized:
because we need to prove properties about stores which differ only on
a set of locations, it is convenient to encode stores and location
sets in a slightly unusual way: location sets are represented as lists
of bits, and stores are represented as lists of value options; in both
representations the $n$th list element corresponds to the $n$th
location. This makes it easy to prove the necessary lemmas by parallel
induction over the lists.

The complete Twelf code can be found in \appref{twelf}

\section{Implementation Strategies}
\label{sec:implementation}
\label{sec:imp}

The dynamic semantics of \aml (\secref{language}) does not translate
directly to an algorithm, not to mention an efficient
one.\footnote{Since our theorems and lemmas concern given derivations
  (not the problem finding them, this does not constitute a problem
  for our results.}  In particular, an algorithm consistent with the
semantics must specify an oracle and a way to allocate locations to
ensure that all locations allocated in a trace are unique.  Strategies
for implementing the semantics beyond the scope of this paper but we
briefly describe a conservative strategy for implementation.  The
strategy ensures that
\begin{enumerate}
\item each allocated location is fresh (i.e., is not contained in the
  memory)
\item the oracle returns only traces currently residing in the memory,
\item the oracle never returns a trace more than once, and
\item the oracle performs function comparisons by using tag equality.
\end{enumerate}

The first two conditions together guarantee uniqueness of allocated
locations.  The third condition guarantees that no location can appear
in the execution trace more than once by limiting the oracle from ever
returning the same trace multiple times.  This condition is
conservative, because it is possible that the parts of a trace
returned by the oracle are thrown away (become unused) during change
propagation.  This strategy can be relaxed by allowing the
change-propagation algorithm to return unused traces to the oracle.
The last condition enables implementing oracle queries by comparing
functions and their arguments by using tag equality.  Since in the
semantics, the oracle is non-deterministic, this implementation
strategy is consistent with the semantics.  The conservative strategy
can be implemented in such a way that the total space consumption is
no more than that of a from-scratch run.  Such an implementation has
been described and evaluated elsewhere~\cite{AcarBlBlHaTa09} and has
formed the basis for subsequent larger-scale
implementations~\cite{HammerAcCh09,Ley-WildFlAc08}.

\section{Related Work}

The related work that this paper directly builds on have been
discussed in the rest of the paper.  Here we briefly discuss other
related on incremental computation and the impact of the result
presented in this paper on follow-up work on self-adjusting
computation.

The term ``incremental computation'' broadly refers to techniques for
allowing computations to respond automatically to changes to their
data.  Motivated by the copious applications where such dynamically
changing data arise, researchers have proposed numerous approaches to
incremental computation.  The most effective techniques are based on
static dependence graphs~\cite{DemersReTe81},
memoization~\cite{PughTe89}, and partial
evaluation~\cite{FieldTe90,SundareshHu91}

Dependence graphs record the dependencies between data in a
computation and rely on a change-propagation algorithm to update the
computation when the input is modified
(e.g.,~\cite{DemersReTe81,Hoover87}).  Dependence graphs are effective
in some applications, e.g., syntax-directed computations but are not
general-purpose because change propagation does not update the
dependence structure.  Memoization (also called function caching)
(e.g.,~\cite{PughTe89, AbadiLaLe96,HeydonLeYu00}) applies to any
purely functional program and therefore is more broadly applicable
than static dependence graphs.  This classic idea dating back to the
late 1950's~\cite{Bellman57,McCarthy63,Michie68} yields efficient
incremental computations when executions of a program with similar
inputs perform similar function calls.  It turns out, however, that
even a small input modifications can prevent reuse via memoization,
e.g., when they affect computations deep in the call
tree~\cite{AcarBlBlHaTa09}.  Partial evaluation based
approaches~\cite{SundareshHu91,FieldTe90} require the user to fix a
partition of the input and specialize the program to speedup
modifications to unfixed part faster.  The main limitation of this
approach is that it allows input modifications only within a
predetermined partition.

The semantics proposed here achieve efficient incremental computation
by integrating a previous generalization of dependence graphs that
allow change propagation to modify the dependence
structure~\cite{AcarBlHa06} with memoization.  Specifically it permits
change propagation algorithm to re-use computations, even after the
computation state is modified via mutations to memory.  In contrast,
conventional memoization permits re-use of the (unchanged) results of
computations.  

The presented semantics forms the foundation for nearly all the
followup work on self-adjusting computation.  After its publication as
a conference paper, the semantics have been realized as a Standard ML
library~\cite{AcarBlBlHaTa09} and generalized to support imperative
references~\cite{AcarAhBl08}.  These results have then led to the
development of the CEAL~\cite{HammerAcCh09} and Delta ML languages,
which provide direct language support for self-adjusting
computation~\cite{Ley-WildFlAc08}.  A relatively broad set of
applications of the proposed techniques have also been investigated,
including simpler computational benchmarks, more sophisticated
applications in computational geometry and machine learning
(e.g.,~\cite{AcarIhMeSu07,AcarCoHuTu10}).  These applications show
that the proposed approach can provide asymptotically optimal updates
in theory while also delivering massive speedups in practice.  In some
cases, the techniques have enabled us to solve open problems that
resisted traditional approaches.

\section{Conclusion}

We present general semantics for integrating memoization and change
propagation where memoization is modeled as a non-deterministic
oracle, and computation re-use is possible in the presence of
mutation.  Mutations arise for two reasons.  First the semantics
permit the store to be modified between two runs while allowing
computations to be re-used between two such runs---this models dynamic
data changes.  Second, the techniques for change propagation mutate
the store by selectively re-executing pieces of the first run to
derive the second run.  The key idea behind the semantics is to enable
re-using of computations themselves by adapting re-used computations
to mutations via recursive applications of change propagation.  Our
main theorem shows that the semantics are consistent with
deterministic, purely functional programming.  By giving a general,
oracle-based semantics for combining memoization and change
propagation, we cover a variety of possible techniques for
implementing self-adjusting-computation.  By proving the semantics
correct with minimal assumptions, we identify the properties that
correct implementations must satisfy.  The results reported in this
work laid out the formal foundation on which other work on
self-adjusting computation has built on. Indeed, the semantics have
been subsequently generalized to imperative programming constructs and
been adapted and realized in strongly typed functional language as
well as procedural, weakly typed languages.

\bibliographystyle{plain}
%\small
\bibliography{main}

\clearpage
\begin{small}
\appendix

\section{The Complete Twelf Proof}
\label{app:twelf} 
\begin{verbatim}
%%%%%%%%%%%%%%%%%%%%%%%%%%%%%%%%%%%%%%%%%%%%%%%%%%%%%%%%%%%%%%%%%%%%%%
%% proof.twelf
%%
%% This file contains the complete Twelf code for the consistency 
%% and correctness proofs for the AML semantics described in  
%% "A Consistent Semantics of Self-Adjusting Computation"
%% by U. A. Acar, M. Blume, J. Donham.
%%
%%%%%%%%%%%%%%%%%%%%%%%%%%%%%%%%%%%%%%%%%%%%%%%%%%%%%%%%%%%%%%%%%%%%%%

%%%%%%%%%%%%%%%%%%%%%%%%%%%%%%%%%%%%%%%%%%%%%%%%%%%%%%%%%%%%%%%%%%%%%%
%% false.elf
%%%%%%%%%%%%%%%%%%%%%%%%%%%%%%%%%%%%%%%%%%%%%%%%%%%%%%%%%%%%%%%%%%%%%%
%% The uninhabited type, indicating a contradiction.
false : type.

%%%%%%%%%%%%%%%%%%%%%%%%%%%%%%%%%%%%%%%%%%%%%%%%%%%%%%%%%%%%%%%%%%%%%%
%% nat.elf
%%%%%%%%%%%%%%%%%%%%%%%%%%%%%%%%%%%%%%%%%%%%%%%%%%%%%%%%%%%%%%%%%%%%%%
%% Natural numbers.

nat	: type. %name nat _N.

z	: nat.
s	: nat -> nat.

nat-eq	: nat -> nat -> type.

nat-eq_	: nat-eq N N.

leq	: nat -> nat -> type.

leq-z	: leq z _.
leq-s	: leq (s N1) (s N2)
	   <- leq N1 N2.

sum	: nat -> nat -> nat ->  type.
%mode sum +X +Y -Z.

sum-z	: sum z N N.
sum-s	: sum (s N1) N2 (s N3)
	   <- sum N1 N2 N3.

%worlds () (sum _ _ _).
%total X (sum X _ _).
%reduces Y <= Z (sum _ Y Z).

%%%%%%%%%%%%%%%%%%%%%%%%%%%%%%%%%%%%%%%%%%%%%%%%%%%%%%%%%%%%%%%%%%%%%%
%% syntax.elf
%%%%%%%%%%%%%%%%%%%%%%%%%%%%%%%%%%%%%%%%%%%%%%%%%%%%%%%%%%%%%%%%%%%%%%
%% Locations are just indices into a store.

loc	: type. %name loc _L.

loc-z	: loc.
loc-s	: loc -> loc.

loc-neq		: loc -> loc -> type.

loc-neq-nil1	: loc-neq loc-z (loc-s L).
loc-neq-nil2	: loc-neq (loc-s L) loc-z.
loc-neq-cons	: loc-neq (loc-s L1) (loc-s L2)
		   <- loc-neq L1 L2.

%% Syntax of AML.

val	: type. %name val _V.
es	: type. %name es _Es.
ec	: type. %name ec _Ec.

val-emp : val.
val-nat	: nat -> val.
val-loc	: loc -> val.
val-pr	: val -> val -> val.
val-inl	: val -> val.
val-inr	: val -> val.
val-fns	: (val -> val -> es) -> val.
val-fnc	: (val -> val -> ec) -> val.

es-val	: val -> es.
es-plus	: val -> val -> es.
es-mod	: ec -> es.
es-memo	: es -> es.
es-app	: val -> val -> es.
es-let	: es -> (val -> es) -> es.
es-letp	: val -> (val -> val -> es) -> es.
es-case	: val -> (val -> es) -> (val -> es) -> es.

ec-wr	: val -> ec.
ec-read	: val -> (val -> ec) -> ec.
ec-memo	: ec -> ec.
ec-app	: val -> val -> ec.
ec-let	: es -> (val -> ec) -> ec.
ec-letp	: val -> (val -> val -> ec) -> ec.
ec-case	: val -> (val -> ec) -> (val -> ec) -> ec.

val-eq	: val -> val -> type.

val-eq_ : val-eq V V.

val-neq	: val -> val -> type.

es-eq	: es -> es -> type.

es-eq_	: es-eq Es Es.

ec-eq	: ec -> ec -> type.

ec-eq_	: ec-eq Ec Ec.

var	: val -> type.

%block val-block : block {v : val}.
%block var-block : block {v : val} {d : var v}.


%%%%%%%%%%%%%%%%%%%%%%%%%%%%%%%%%%%%%%%%%%%%%%%%%%%%%%%%%%%%%%%%%%%%%%
%% locset.elf
%%%%%%%%%%%%%%%%%%%%%%%%%%%%%%%%%%%%%%%%%%%%%%%%%%%%%%%%%%%%%%%%%%%%%%

%% Sets of locations. We represent them as lists of bits; lists which
%% differ only by trailing false bits are equivalent.

loc-state : type.
loc-present : loc-state.
loc-absent : loc-state.

loc-or : loc-state -> loc-state -> loc-state -> type.
loc-or-aa : loc-or loc-absent loc-absent loc-absent.
loc-or-px : loc-or loc-present _ loc-present.
loc-or-xp : loc-or _ loc-present loc-present.

ls	: type. %name ls _X.

ls-nil     : ls.
ls-cons     : loc-state -> ls -> ls.

%% check for empty set
ls-empty : ls -> type.
ls-empty-n : ls-empty ls-nil.
ls-empty-a : ls-empty (ls-cons loc-absent X)
	      <- ls-empty X.

%% set equality
ls-eq : ls -> ls -> type.
ls-eq-nx : ls-eq ls-nil X
	    <- ls-empty X.
ls-eq-xn : ls-eq X ls-nil
	    <- ls-empty X.
ls-eq-cc : ls-eq (ls-cons P X1) (ls-cons P X2)
	    <- ls-eq X1 X2.

%% representation identity
ls-id : ls -> ls -> type.
ls-id_ : ls-id X X.

%% X_1 \subseteq X_2
ls-subeq : ls -> ls -> type.
ls-subeq-nx : ls-subeq ls-nil _.
ls-subeq-xn : ls-subeq X ls-nil
	       <- ls-empty X.
ls-subeq-ax : ls-subeq (ls-cons loc-absent X1) (ls-cons _ X2)
	       <- ls-subeq X1 X2.
ls-subeq-pp : ls-subeq (ls-cons loc-present X1) (ls-cons loc-present X2)
	       <- ls-subeq X1 X2.

%% X_1 \cup X_2
ls-union : ls -> ls -> ls -> type.
ls-un-nx : ls-union ls-nil X X.
ls-un-xn : ls-union X ls-nil X.
ls-un-cc : ls-union (ls-cons P1 X1) (ls-cons P2 X2) (ls-cons P X)
	    <- loc-or P1 P2 P
	    <- ls-union X1 X2 X.

%% X_1 \cap X_2 = 0
ls-disjoint : ls -> ls -> type.
ls-dj-nx : ls-disjoint ls-nil _.
ls-dj-xn : ls-disjoint _ ls-nil.
ls-dj-ac : ls-disjoint (ls-cons loc-absent X) (ls-cons _ X')
	    <- ls-disjoint X X'.
ls-dj-ca : ls-disjoint (ls-cons _ X) (ls-cons loc-absent X')
	    <- ls-disjoint X X'.

%% Set of a single location
ls-sing : loc -> ls -> type.
ls-sing-z : ls-sing loc-z (ls-cons loc-present ls-nil).
ls-sing-s : ls-sing (loc-s L) (ls-cons loc-absent S)
	     <- ls-sing L S.

%%%%%%%%%%%%%%%%%%%%%%%%%%%%%%%%%%%%%%%%%%%%%%%%%%%%%%%%%%%%%%%%%%%%%%
%% store.elf
%%%%%%%%%%%%%%%%%%%%%%%%%%%%%%%%%%%%%%%%%%%%%%%%%%%%%%%%%%%%%%%%%%%%%%

%% Stores mapping locations to values. We represent them as lists of
%% value options, where the i'th element of the list is the value of
%% location i in the store (or sv-free if the location is
%% undefined). As with location sets, stores differing only by trailing
%% sv-free's are equivalent.

%% We choose the bitwise representations because it makes the lemmas
%% of interest easier to prove; they are generally just an induction
%% over the bits.

st	: type. %name st _S.

sv : type. %name sv _SV.   %% store value: either free or a value

sv-free : sv.
sv-val : val -> sv.

sv-eq : sv -> sv -> type.
sv-eq_ : sv-eq SV SV.

st-nil	: st.
st-cons: sv -> st -> st.

%% Are all locations empty?
st-empty : st -> type.
st-empty-n : st-empty st-nil.
st-empty-e : st-empty (st-cons sv-free S)
	      <- st-empty S.

%% Store equality.  (Could we get away with syntactic equality? Probably.)
st-eq	: st -> st -> type.
st-eq-nx : st-eq st-nil S
	    <- st-empty S.
st-eq-xn : st-eq S st-nil
	    <- st-empty S.
st-eq-cc : st-eq (st-cons SV1 S1) (st-cons SV2 S2)
	    <- sv-eq SV1 SV2
	    <- st-eq S1 S2.

%% \sigma [l \leftarrow v]
st-update	: st -> loc -> val -> st -> type.

st-up-nz : st-update st-nil loc-z V (st-cons (sv-val V) st-nil).
st-up-cz : st-update (st-cons _ S) loc-z V (st-cons (sv-val V) S).
st-up-ns : st-update st-nil (loc-s L) V (st-cons sv-free S)
	    <- st-update st-nil L V S.
st-up-cs : st-update (st-cons SV S) (loc-s L) V (st-cons SV S')
	    <- st-update S L V S'.

%% \sigma(l)
st-lookup : st -> loc -> val -> type.
st-lo-z : st-lookup (st-cons (sv-val V) _) loc-z V.
st-lo-s : st-lookup (st-cons _ S) (loc-s L) V
	   <- st-lookup S L V.

%% st-sqsubeq-ex S1 X S2 holds if for any location L allocated in S1 with
%%        value V, either L is in X or S2 has value V at location L.
%%
%% This is rather painful because of the treatment of ls-nil and st-nil
%% in the 2nd and 3rd arguments, respectively.
st-sqsubeq-ex : st -> ls -> st -> type.
st-ssee-nxx : st-sqsubeq-ex st-nil _ _.
st-ssee-fnn : st-sqsubeq-ex (st-cons sv-free S1') ls-nil st-nil
	       <- st-sqsubeq-ex S1' ls-nil st-nil.
st-ssee-fcn : st-sqsubeq-ex (st-cons sv-free S1') (ls-cons _ X') st-nil
	       <- st-sqsubeq-ex S1' X' st-nil.
st-ssee-fnc : st-sqsubeq-ex (st-cons sv-free S1') ls-nil (st-cons _ S2')
	       <- st-sqsubeq-ex S1' ls-nil S2'.
st-ssee-fcc : st-sqsubeq-ex (st-cons sv-free S1') (ls-cons _ X') (st-cons _ S2')
	       <- st-sqsubeq-ex S1' X' S2'.
st-ssee-vnv : st-sqsubeq-ex
                  (st-cons (sv-val V1) S1') ls-nil (st-cons (sv-val V2) S2')
	       <- val-eq V1 V2
	       <- st-sqsubeq-ex S1' ls-nil S2'.
st-ssee-vav : st-sqsubeq-ex
                  (st-cons (sv-val V1) S1') (ls-cons loc-absent X')
	          (st-cons (sv-val V2) S2')
	       <- val-eq V1 V2
	       <- st-sqsubeq-ex S1' X' S2'.
st-ssee-cpn : st-sqsubeq-ex (st-cons _ S1') (ls-cons loc-present X') st-nil
	       <- st-sqsubeq-ex S1' X' st-nil.
st-ssee-cpc : st-sqsubeq-ex
	          (st-cons _ S1') (ls-cons loc-present X')
	          (st-cons _ S2')
	       <- st-sqsubeq-ex S1' X' S2'.


%%%%%%%%%%%%%%%%%%%%%%%%%%%%%%%%%%%%%%%%%%%%%%%%%%%%%%%%%%%%%%%%%%%%%%
%% trace.elf
%%%%%%%%%%%%%%%%%%%%%%%%%%%%%%%%%%%%%%%%%%%%%%%%%%%%%%%%%%%%%%%%%%%%%%
%% Evaluation traces, and their allocated locations.

trs	: type. %name trs _Ts.
trc	: type. %name trc _Tc.

trs-nil	: trs.
trs-mod	: loc -> trc -> trs.
trs-let	: trs -> trs -> trs.

trc-wr	: val -> trc.
trc-let	: trs -> trc -> trc.
trc-rd	: loc -> val -> (val -> ec) -> trc -> trc.

trs-gen		: trs -> ls -> type.
trc-gen		: trc -> ls -> type.

trs-gen-nil	: trs-gen trs-nil ls-nil.
trs-gen-mod	: trs-gen (trs-mod L Tc) X1+X2
		   <- trc-gen Tc X1
		   <- ls-sing L X2
		   <- ls-union X1 X2 X1+X2.
trs-gen-let	: trs-gen (trs-let Ts1 Ts2) X
		   <- trs-gen Ts1 X1
		   <- trs-gen Ts2 X2
		   <- ls-union X1 X2 X.

trc-gen-wr	: trc-gen (trc-wr V) ls-nil.
trc-gen-let	: trc-gen (trc-let Ts1 Tc2) X
		   <- trs-gen Ts1 X1
		   <- trc-gen Tc2 X2
		   <- ls-union X1 X2 X.
trc-gen-rd	: trc-gen (trc-rd L V Ec Tc) X
		   <- trc-gen Tc X.

%%%%%%%%%%%%%%%%%%%%%%%%%%%%%%%%%%%%%%%%%%%%%%%%%%%%%%%%%%%%%%%%%%%%%%
%% wf-ex.elf
%%%%%%%%%%%%%%%%%%%%%%%%%%%%%%%%%%%%%%%%%%%%%%%%%%%%%%%%%%%%%%%%%%%%%%
%% Well-formed expressions (with lifts and reachable locations)

wf-val	: val -> st -> val -> ls -> type.
wf-es	: es -> st -> es -> ls -> type.
wf-ec	: ec -> st -> ec -> ls -> type.

wf-val-var	: wf-val V S V ls-nil
		   <- var V.

wf-val-emp	: wf-val val-emp S val-emp ls-nil.
wf-val-nat	: wf-val (val-nat N) S (val-nat N) ls-nil.
wf-val-loc	: wf-val (val-loc L) S V' X1+X2
		   <- st-lookup S L V
		   <- wf-val V S V' X1
		   <- ls-sing L X2
		   <- ls-union X1 X2 X1+X2.
wf-val-pr	: wf-val (val-pr V1 V2) S (val-pr V1' V2') X1+X2
		   <- wf-val V1 S V1' X1
		   <- wf-val V2 S V2' X2
		   <- ls-union X1 X2 X1+X2.
wf-val-inl	: wf-val (val-inl V) S (val-inl V') X
		   <- wf-val V S V' X.
wf-val-inr	: wf-val (val-inr V) S (val-inr V') X
		   <- wf-val V S V' X.
wf-val-fns	: wf-val (val-fns Es) S (val-fns Es') X
		   <- ({v1}{d1 : var v1}
		       {v2}{d2 : var v2}
			 wf-es (Es v1 v2) S (Es' v1 v2) X).
wf-val-fnc	: wf-val (val-fnc Ec) S (val-fnc Ec') X
		   <- ({v1}{d1 : var v1}
                       {v2}{d2 : var v2}
			 wf-ec (Ec v1 v2) S (Ec' v1 v2) X).

wf-es-val	: wf-es (es-val V) S (es-val V') X
		   <- wf-val V S V' X.
wf-es-plus	: wf-es (es-plus V1 V2) S (es-plus V1' V2') X
		   <- wf-val V1 S V1' X1
		   <- wf-val V2 S V2' X2
		   <- ls-union X1 X2 X.
wf-es-mod	: wf-es (es-mod Ec) S (es-mod Ec') X
		   <- wf-ec Ec S Ec' X.
wf-es-app	: wf-es (es-app V1 V2) S (es-app V1' V2') X
		   <- wf-val V1 S V1' X1
		   <- wf-val V2 S V2' X2
		   <- ls-union X1 X2 X.
wf-es-let	: wf-es (es-let Es1 Es2) S (es-let Es1' Es2') X
		   <- wf-es Es1 S Es1' X1
		   <- ({v}{d : var v}
			 wf-es (Es2 v) S (Es2' v) X2)
		   <- ls-union X1 X2 X.
wf-es-letp	: wf-es (es-letp V Es) S (es-letp V' Es') X
		   <- wf-val V S V' X1
		   <- ({v1}{d1 : var v1}
		       {v2}{d2 : var v2}
			 wf-es (Es v1 v2) S (Es' v1 v2) X2)
		   <- ls-union X1 X2 X.
wf-es-case	: wf-es (es-case V Es1 Es2) S (es-case V' Es1' Es2') X
		   <- wf-val V S V' X0
		   <- ({v}{d : var v}
			 wf-es (Es1 v) S (Es1' v) X1)
		   <- ({v}{d : var v}
			 wf-es (Es2 v) S (Es2' v) X2)
		   <- ls-union X1 X2 X12
		   <- ls-union X12 X0 X.
wf-es-memo	: wf-es (es-memo Es) S (es-memo Es') X
		   <- wf-es Es S Es' X.

wf-ec-wr	: wf-ec (ec-wr V) S (ec-wr V') X
		   <- wf-val V S V' X.
wf-ec-read	: wf-ec (ec-read V Ec) S (ec-read V' Ec') X
		   <- wf-val V S V' X1
		   <- ({v}{d : var v}
			 wf-ec (Ec v) S (Ec' v) X2)
		   <- ls-union X1 X2 X.
wf-ec-app	: wf-ec (ec-app V1 V2) S (ec-app V1' V2') X
		   <- wf-val V1 S V1' X1
		   <- wf-val V2 S V2' X2
		   <- ls-union X1 X2 X.
wf-ec-let	: wf-ec (ec-let Es1 Ec2) S (ec-let Es1' Ec2') X
		   <- wf-es Es1 S Es1' X1
		   <- ({v}{d : var v}
			 wf-ec (Ec2 v) S (Ec2' v) X2)
		   <- ls-union X1 X2 X.
wf-ec-letp	: wf-ec (ec-letp V Ec) S (ec-letp V' Ec') X
		   <- wf-val V S V' X1
		   <- ({v1}{d1 : var v1}
		       {v2}{d2 : var v2}
			 wf-ec (Ec v1 v2) S (Ec' v1 v2) X2)
		   <- ls-union X1 X2 X.
wf-ec-case	: wf-ec (ec-case V Ec1 Ec2) S (ec-case V' Ec1' Ec2') X
		   <- wf-val V S V' X0
		   <- ({v}{d : var v}
			 wf-ec (Ec1 v) S (Ec1' v) X1)
		   <- ({v}{d : var v}
			 wf-ec (Ec2 v) S (Ec2' v) X2)
		   <- ls-union X1 X2 X12
		   <- ls-union X12 X0 X.
wf-ec-memo	: wf-ec (ec-memo Ec) S (ec-memo Ec') X
		   <- wf-ec Ec S Ec' X.


%%%%%%%%%%%%%%%%%%%%%%%%%%%%%%%%%%%%%%%%%%%%%%%%%%%%%%%%%%%%%%%%%%%%%%
%% eval.elf
%%%%%%%%%%%%%%%%%%%%%%%%%%%%%%%%%%%%%%%%%%%%%%%%%%%%%%%%%%%%%%%%%%%%%%
%% General, well-formed, and clean evaluations.

evals		: st -> es -> val -> st -> trs -> type.
evalc		: st -> loc -> ec -> st -> trc -> type.

wf-evals	: es -> ls -> ls -> evals _ _ _ _ _ -> type.
wf-evalc	: ec -> ls -> ls -> ls -> evalc _ _ _ _ _ -> type.

wf-evals_	: wf-evals Es' R G (Devals : evals S Es V S' Ts)
		   <- wf-es Es S Es' R
		   <- trs-gen Ts G
		   <- ls-disjoint R G.

wf-evalc_	: wf-evalc Ec' R G X (Devalc : evalc S L Ec S' Tc)
		   <- wf-ec Ec S Ec' R
		   <- trc-gen Tc G
		   <- ls-disjoint R G
		   <- ls-sing L X
		   <- ls-disjoint X R
		   <- ls-disjoint X G.

cps		: st -> trs -> st -> trs -> type.
cpc		: st -> loc -> trc -> st -> trc -> type.

evals-val	: evals S (es-val V) V S trs-nil.
evals-plus	: evals S (es-plus (val-nat N1) (val-nat N2)) (val-nat N3) S trs-nil
		   <- sum N1 N2 N3.
evals-mod	: evals S (es-mod Ec) (val-loc L) S' (trs-mod L Tc)
		   <- evalc S L Ec S' Tc
		   <- trc-gen Tc G
		   <- ls-sing L X
		   <- ls-disjoint X G.
evals-memo-miss	: evals S (es-memo Es) V S' Ts
		   <- evals S Es V S' Ts.

%% can we mix backward arrows with Pi's?
evals-memo-hit	:     cps S Ts1 S' Ts
		   -> {Devals : evals S1 Es V S1' Ts1}
		      wf-evals Es' R G Devals
		   -> evals S (es-memo Es) V S' Ts.

evals-app	: evals S (es-app (val-fns Es) V2) V S' Ts
		   <- evals S (Es (val-fns Es) V2) V S' Ts.
evals-let	: evals S (es-let Es1 Es2) V2 S2 (trs-let Ts1 Ts2)
		   <- evals S Es1 V1 S1 Ts1
		   <- evals S1 (Es2 V1) V2 S2 Ts2
		   <- trs-gen Ts1 G1
		   <- trs-gen Ts2 G2
		   <- ls-disjoint G1 G2.
evals-letp	: evals S (es-letp (val-pr V1 V2) Es) V S' Ts
		   <- evals S (Es V1 V2) V S' Ts.
evals-case-inl	: evals S (es-case (val-inl V) Es1 Es2) V' S' Ts
		   <- evals S (Es1 V) V' S' Ts.
evals-case-inr	: evals S (es-case (val-inr V) Es1 Es2) V' S' Ts
		   <- evals S (Es2 V) V' S' Ts.

evalc-write	: evalc S L (ec-wr V) S' (trc-wr V)
		   <- st-update S L V S'.
evalc-read	: evalc S L' (ec-read (val-loc L) Ec) S' (trc-rd L V Ec Tc)
		   <- st-lookup S L V
		   <- evalc S L' (Ec V) S' Tc.
evalc-memo-miss	: evalc S L (ec-memo Ec) S' Tc
		   <- evalc S L Ec S' Tc.

%% can we mix backward arrows with Pi's?
evalc-memo-hit	:     cpc S L Tc1 S' Tc
		   -> {Devalc : evalc S1 L Ec S1' Tc1}
		      wf-evalc Ec' R G X Devalc
		   -> evalc S L (ec-memo Ec) S' Tc.

evalc-app	: evalc S L (ec-app (val-fnc Ec) V2) S' Tc
		   <- evalc S L (Ec (val-fnc Ec) V2) S' Tc.
evalc-let	: evalc S L (ec-let Es1 Ec2) S2 (trc-let Ts1 Tc2)
		   <- evals S Es1 V S1 Ts1
		   <- evalc S1 L (Ec2 V) S2 Tc2
		   <- trs-gen Ts1 G1
		   <- trc-gen Tc2 G2
		   <- ls-disjoint G1 G2.
evalc-letp	: evalc S L (ec-letp (val-pr V1 V2) Ec) S' Tc
		   <- evalc S L (Ec V1 V2) S' Tc.
evalc-case-inl	: evalc S L (ec-case (val-inl V) Ec1 Ec2) S' Tc
		   <- evalc S L (Ec1 V) S' Tc.
evalc-case-inr	: evalc S L (ec-case (val-inr V) Ec1 Ec2) S' Tc
		   <- evalc S L (Ec2 V) S' Tc.

cps-nil		: cps S trs-nil S trs-nil.
cps-mod		: cps S (trs-mod L Tc) S' (trs-mod L Tc')
		   <- cpc S L Tc S' Tc'
		   <- trc-gen Tc' G
		   <- ls-sing L X
		   <- ls-disjoint X G.
cps-let		: cps S (trs-let Ts1 Ts2) S'' (trs-let Ts1' Ts2')
		   <- cps S Ts1 S' Ts1'
		   <- cps S' Ts2 S'' Ts2'
		   <- trs-gen Ts1' G1
		   <- trs-gen Ts2' G2
		   <- ls-disjoint G1 G2.

cpc-write	: cpc S L (trc-wr V) S' (trc-wr V)
		   <- st-update S L V S'.
cpc-let		: cpc S L' (trc-let Ts1 Tc2) S'' (trc-let Ts1' Tc2')
		   <- cps S Ts1 S' Ts1'
		   <- cpc S' L' Tc2 S'' Tc2'
		   <- trs-gen Ts1' G1
		   <- trc-gen Tc2' G2
		   <- ls-disjoint G1 G2.
cpc-read/noch	: cpc S L (trc-rd L' V Ec Tc) S' (trc-rd L' V Ec Tc')
		   <- st-lookup S L' V
		   <- cpc S L Tc S' Tc'.
cpc-read/ch	: cpc S L (trc-rd L' V Ec Tc) S' (trc-rd L' V' Ec Tc')
		   <- st-lookup S L' V'
		   <- val-neq V V'
		   <- evalc S L (Ec V') S' Tc'.

%%

cln-evals	: evals _ _ _ _ _ -> type.
cln-evalc	: evalc _ _ _ _ _ -> type.

cln-evals-val	: cln-evals evals-val.
cln-evals-plus	: cln-evals (evals-plus _).
cln-evals-mod	: cln-evals (evals-mod _ _ _ D)
		   <- cln-evalc D.
cln-evals-miss	: cln-evals (evals-memo-miss D)
		   <- cln-evals D.
cln-evals-app	: cln-evals (evals-app D)
		   <- cln-evals D.
cln-evals-let	: cln-evals (evals-let _ _ _ D2 D1)
		   <- cln-evals D1
		   <- cln-evals D2.
cln-evals-letp	: cln-evals (evals-letp D)
		   <- cln-evals D.
cln-evals-inl	: cln-evals (evals-case-inl D)
		   <- cln-evals D.
cln-evals-inr	: cln-evals (evals-case-inr D)
		   <- cln-evals D.

cln-evalc-write	: cln-evalc (evalc-write _).
cln-evalc-read	: cln-evalc (evalc-read D _)
		   <- cln-evalc D.
cln-evalc-miss	: cln-evalc (evalc-memo-miss D)
		   <- cln-evalc D.
cln-evalc-app	: cln-evalc (evalc-app D)
		   <- cln-evalc D.
cln-evalc-let	: cln-evalc (evalc-let _ _ _ D2 D1)
		   <- cln-evals D1
		   <- cln-evalc D2.
cln-evalc-letp	: cln-evalc (evalc-letp D)
		   <- cln-evalc D.
cln-evalc-inl	: cln-evalc (evalc-case-inl D)
		   <- cln-evalc D.
cln-evalc-inr	: cln-evalc (evalc-case-inr D)
		   <- cln-evalc D.


\end{verbatim}

\clearpage


\end{small}

\end{document}